\documentclass{elsart}

\usepackage{amssymb,pictex,graphicx}

\journal{Journal of Physics A}
 
\begin{document}
\begin{frontmatter}

\title{Vacuum Stress and Closed Paths \\
 in Rectangles, Pistons, and Pistols}

 \author[tamumath,tamuphys]{S. A. Fulling\corauthref{cor}},
 \corauth[cor]{Corresponding author.}
 \ead{fulling@math.tamu.edu} 
 \author[tulane]{L. Kaplan},
 \author[baylor]{K. Kirsten},
  \author[tamuphys]{Z. H. Liu},
 \author[ou]{K. A. Milton}

  \address[tamumath]{Department of Mathematics, Texas A\&M 
University, College Station, TX, 77843-3368 USA}

 \address[tamuphys]{Department of Physics, Texas A\&M University,
 College Station, TX,\break\ 77843-4242 USA} 

  \address[tulane]{Department of Physics, Tulane University,
      New Orleans, LA, 70118 USA}

  \address[baylor]{Department of Mathematics, Baylor University,
 Waco, TX, 76798-7328 USA}

  \address[ou]{H. L. Dodge Department of Physics and Astronomy, 
 University of Oklahoma,
 Norman, OK, 73019-2061 USA}

\begin{abstract}
Rectangular cavities are solvable models that nevertheless touch on 
many of the controversial or mysterious aspects of the vacuum 
energy of quantum fields.
 This paper  is a thorough study of the two-dimensional scalar 
field in a rectangle by the method of images, or closed  classical 
(or optical) paths, which is exact in this case. 
 For each point $\mathbf{r}$ and each specularly reflecting path 
beginning and ending at~$\mathbf{r}$, we provide formulas for all 
components of the stress tensor $T_{\mu\nu}(\mathbf{r})$,
 for all values of the curvature coupling constant $\xi$ and 
 all values of an ultraviolet cutoff parameter. 
  Arbitrary combinations of 
Dirichlet and Neumann conditions on the four sides can be treated.
 The total energy is also investigated, path by path.
These results are used in an attempt to clarify the physical 
reality of the repulsive 
(outward) force on the sides of the box predicted by calculations 
that neglect both boundary divergences and the exterior of the box.
 Previous authors have studied  ``piston'' geometries that avoid 
these problems and have found the force to be attractive.
 We consider a ``pistol'' geometry that comes closer to the 
original problem of a box with a movable lid.  
We find again an attractive force, although its origin and
detailed behavior are somewhat different from the piston case.
However, the pistol (and the piston) model can be criticized for
extending idealized boundary conditions into short distances
where they are physically implausible.  
    Therefore, it is of interest to see whether leaving the 
ultraviolet cutoff finite yields results that are more plausible. 
     We then find that the force depends 
 strongly  on a geometrical parameter;
 it can be made repulsive, but only by forcing that parameter 
into 
the regime where the model is least convincing physically.
          \end{abstract} 
 
 \begin{keyword} 
 Casimir \sep vacuum energy \sep parallelepiped \sep ultraviolet cutoff
 \PACS 03.70.+k \sep 04.20.Cv \sep 11.10.Gh \sep 45.20.Lc
 \end{keyword}

\end{frontmatter}

 \newcommand{\tbar}{\overline{T}}
 \newcommand{\bra}{\left\langle}
 \newcommand{\ket}{\right\rangle}
 \newcommand{\pd}[2]{\frac{\partial #1}{\partial #2}}
 \newcommand{\od}[2]{\frac{d#1}{d#2}}
 \newcommand{\q}{({\scriptstyle\frac14})}
\newcommand{\gfatop}[2]{{\genfrac{}{}{0pt}{1}{#1}{#2}}}


\section{Introduction} \label{sec:recintro}

 \subsection{History and motivation} \label{ssec:motiv}

Rectangular cavities are perhaps the most frequently studied 
geometries in connection with vacuum (Casimir) energy.
 Nevertheless, there are still worthwhile things to say about them.
 Although exactly solvable, they exhibit many  features that 
are subjects of current research and debate in the broader context 
of quantum vacuum energy: 
boundary divergences, corner effects, sometimes surprising signs, 
and sometimes revealing connections with geometry through the 
spectrum of periodic and other closed classical paths (or optical 
rays).

A complete listing of previous literature is impossible, but we 
summarize what we see as the most important historical 
developments.

 Lukosz \cite{Lu} calculated the interior vacuum energy of the 
electromagnetic field in a (3D) perfectly conducting 
parallelepiped, using zeta-function regularization, and predicted a 
repulsive (outward) force for many aspect ratios, including the 
cube. 
 Ambj{\o}rn and Wolfram \cite{AW} extended such calculations to a 
wide variety of dimensions,   fields, and boundary conditions.
 Actor \cite{A} emphasized that divergences in the total energy 
must be understood in terms of the local behavior of the field near 
boundaries and boundary singularities (edges and corners), 
 and calculated the local zeta function for the 3D scalar field.
For earlier, closely related discussions of a rectangular
waveguide and various other systems, see \cite{DowK,DowB}. 
All these works  were done in 
the framework of zeta functions~\cite{El,EORBZ,Kir},
  but in practice, in the special 
case of rectangular cavities, functional equations for Epstein zeta 
functions are used to convert zeta-regularized sums over 
eigenvalues into what are, in effect, zeta-regularized sums over 
classical paths.
(In \cite{RVZ} a transition to an ultraviolet cutoff was also
made at this step.)

 The calculation of vacuum energy via classical paths (also called 
optical rays) \cite{BmC,BD12,JR,SS,MSSV,funorman,JS,SJ12} leads 
naturally to more physical regularizations associated with 
separation of points in the Green functions of the theory.
 For rectangular parallelepipeds such calculations are exact
 (no stationary-phase approximations are required) and reduce to 
the classic method of images.  Although they did not discuss vacuum 
energy, Balian and Bloch \cite{BB3} used the 3D parallelepiped 
as a principal example in their landmark study of the relation 
between periodic orbits and oscillations in the eigenvalue density, 
and their catalog of periodic and other closed orbits is the 
best starting point for a study of the rectangle.
 Hacyan et al.~\cite{HJV} calculated the full stress tensor for the 
electromagnetic field in a box by a Green-function approach;
 along the way, they showed how to reduce the electromagnetic problem
 to scalar fields with mixed Dirichlet and Neumann boundary conditions
 via Hertz potentials. In this connection see also \cite{RS}.
 (Contrary to the impression left by some papers on the global 
problem, a local investigation of electromagnetism cannot be split 
into pure Dirichlet and pure Neumann problems, even when a 
decomposition into TE and TM modes exists.) 

 At this point we should mention the work of Ford and Svaiter 
\cite{FS}, which showed that physically motivated cutoffs could 
convert divergences into finite effects clearly localized near 
boundaries.
 This theme has been repeatedly visited since then
 \cite{GO,systemat,funorman,delta,ines1,ines2}
 and will play a major role in the present work.

 Cavalcanti \cite{Cav} rejuvenated the field  by introducing the 
piston model (for a 2D scalar field), 
 discussed in detail in Sec.~\ref{sec:piston} and illustrated in 
Fig~\ref{fig:piston}.
  (Similar ideas were advanced earlier by Svaiter et 
al.~\cite{Sv}.)
 The motivation for the piston is that the calculation of the 
force on the piston plate is unaffected by either uncompensated 
divergences or unknown forces from the exterior.
 The conclusion of \cite{Cav} is that the force is always attractive 
(inward).
 That paper used both zeta and cutoff regularization, but still 
starting from the eigenvalue spectrum.
 Hertzberg et al.\ \cite{HJ1,HJ2} extended the piston model to 
dimension~3 and to the electromagnetic field, and they analyzed it 
in terms of closed paths (but without the close attention to 
locally defined quantities that we provide here). 
 From that point of view, the repulsive nature of the Lukosz force 
is attributable to a particular type of path moving parallel to the 
plate and producing an energy proportional to the piston 
displacement.  (It is essentially the Casimir energy associated 
with the walls perpendicular to the movable plate.)  But such energy is 
also present in the exterior part of the piston shaft, and 
therefore these paths make no net contribution to the force.  What 
is left of the Lukosz force is attractive. 
 This effect shows up even more clearly in the two-dimensional 
model (Sec.~\ref{sec:piston}).

Rodriguez et al.\ \cite{R1,R2} have made a numerical study of two
conducting rectangular objects in a narrow passage, a model closely 
akin to the pistons and pistols we discuss here.
 They conclude that the distance to the confining walls influences 
the attraction between the blocks, and their analysis makes use of 
the local stress tensor.
 In \cite{Z} that model is approached by the method of closed 
optical paths.

Illuminating though the piston has been, it does not settle the 
original issue of the physical reality of the force calculated by 
Lukosz \cite{Lu} and others.
 The existence of a Casimir-like energy in the exterior part of the 
piston shaft says nothing about what happens when that part of the 
shaft is removed, the plate remaining free to move 
 (see Fig.~\ref{fig:looselid} in  Sec.~\ref{sec:pistol-d}).
 The ``finite part'' of that force is robust, in the sense that all 
reasonable prescriptions for calculating it give the same answer.  
It can be obtained by differentiating the total energy, or by 
integrating the pressure over the movable boundary.
 It can be obtained by zeta-function regularization or by 
ultraviolet cutoffs, and within the latter framework the choice of 
cutoff function dictates the relative sizes of the 
 cutoff-dependent terms but not the structure of the series nor the 
numerical value of the finite term \cite[Appendix B]{Cav}.
 Is the object of this consensus a meaningless number?  One of 
our goals is to investigate to what extent it has physical
significance.

 The opinion expressed in \cite{HJ1} is that 
 ``Without [the piston shaft] (or some open region that allows 
rigid motion of the partition) the Casimir energy of the 
parallelepiped is, in fact, cutoff dependent.
 If the cutoff dependence is somehow ignored, a repulsive force 
\dots\ remains as an artifact.''
 We agree that a correct calculation of the  force on the piston 
must include the effect of the piston shaft,  
and that the net effect is attractive.
 We do not  agree that the repulsive force 
associated with the interior can be dismissed as 
an artifact of naively discarding a divergent term. 
The scenario indicated in Fig.~\ref{fig:looselid}, 
 a box with a movable lid, is a 
 well-defined problem of relative motion of rigid bodies, just as 
much as the piston is; cutoff-dependent energies associated with 
the rigid boundaries cannot affect the force.
The difficulties of analyzing Fig.~\ref{fig:looselid} are, first, 
that the effects of the corners and gaps in that geometry are hard 
to calculate
(see, however, \cite{Luedge,DC,Dowker,Smith,GL} and, on a
different tack, \cite{MPW,MPWconfs}),
 and, second, that the idealized Casimir theory is not 
physically applicable to very small separations of the bodies.
 We evade the first problem by considering another scenario,
 the ``pistol'' (Fig.~\ref{fig:pistol}),
 which should still exhibit the uncompensated Lukosz force on a 
flat boundary. 
     However, we find that the situation is then confounded by a 
strong countervailing attractive force associated with the 
Casimir energy in the narrow gap surrounding the
``bullet''\negthinspace.
Moreover, one now runs up against the second problem, which 
cannot be treated seriously within the limits of our methodology.  
A crude model of a ``real'' boundary can be easily obtained, 
however, by maintaining a finite cutoff of atomic dimensions.
    The result is that the force depends sensitively on how tightly 
the ``bullet'' fits into the ``barrel''\negthinspace.
 If the fit is loose, the Lukosz force is overwhelmed by the 
corresponding force associated with the gap surrounding the 
bullet, and the net force is attractive.
 If the fit is tight, the gap force can be made repulsive or even
 fine-tuned to vanish, as originally hoped; unfortunately,  that 
is the regime in which one is least justified in taking the model 
 seriously.
All we claim is that external forces opposing the Lukosz force are 
model-dependent and might, in principle, be controlled so as to 
demonstrate the existence of the Lukosz force.

 In this paper we consider strictly the two-dimensional scalar 
field,
usually with Dirichlet boundary conditions,
 although we sometimes lapse into the three-dimensional 
electromagnetic terminology (such as ``conductor'') for conceptual 
discussions.
 It is intended that  three-dimensional generalizations will be 
presented elsewhere~\cite{Liu}.
 Sec.~\ref{sec:local} 
 presents a thorough analysis, by means of classical paths, of all 
components of the stress (energy-momentum) tensor in 
a rectangle.
 Sec.~\ref{sec:global} does the same for the energy and also the 
 pressure and force on one side. 
 Contributions are recorded for each path (or class of similar 
paths) separately, with comments on their physical or geometrical 
significance to the extent that we can discern it.
 The results are stated for all values of the 
curvature, or conformal, coupling constant, $\xi$
  (see (\ref{lagrangian}) and (\ref{stresst})),
  and all values of the parameter in an 
exponential ultraviolet cutoff.
 For the most part, they are stated for any combination of 
Dirichlet and Neumann conditions on the four sides of the box.
 A brief account of this part of the work was published 
previously~\cite{leipzig}, along with evidence that the 
gravitational effects of boundaries in the 
 ``renormalized'' theory 
without cutoff can be understood (and believed) as the 
distributional limit of the predictions of the cutoff theory, 
thereby providing a true renormalization.  
 In the rest of the paper we restrict to the Dirichlet condition.
 The piston is reviewed from our point of view in 
Sec.~\ref{sec:piston}.
Sec.~\ref{sec:pistol-d} introduces the pistol model and treats it 
in the Dirichlet theory.
Sec.~\ref{sec:pistol-c}  investigates the pistol with a finite 
cutoff.

{\sl A remark on terminology:\/}
Many authors, including some of ourselves on previous occasions 
(e.g., \cite{systemat}), use the term ``renormalized energy'' to 
refer to  the finite part of a regularized energy when the 
latter is expanded as a series in the cutoff parameter.
Strictly speaking, ``renormalization'' refers to the process  
of obtaining  physically observable probability amplitudes by 
absorbing suitable divergent and finite contributions into 
redefinitions of physical parameters (couplings, masses, etc.)\
appearing in the bare Lagrangian.
  Ideally, 
all renormalizations in the first sense should either be 
associated with renormalizations in the second sense or be 
justified by cancellations of divergent terms coming from 
different sources.
Yet in the absence of a completed theory, one must often talk 
about renormalization in the first sense without having  an 
obvious counterterm or cancellation, and there seems to be no 
convenient substitute terminology.
Much of our work in this paper has to do, in fact, with 
exhibiting cancellations, and \cite{leipzig} and its planned 
sequels have to do with gravitational counterterms.  
When we use ``renormalization'' or ``renormalized'' in the first 
sense, we have always either put the word in quotation marks or 
accompanied it by the word ``naive''\negthinspace, depending on 
context.

 \subsection{Basic formalism} \label{ssec:formalism}

 We are concerned here with the massless scalar wave equation
\begin{equation}
  \pd{^2\phi}{(x^0)^2} = \nabla^2\phi
 \label{fieldeq}\end{equation}
 in a cavity $\Omega$ together with a Dirichlet ($\phi=0$)
 or Neumann ($\hat\mathbf{n}\cdot \nabla\phi=0$) condition 
on each part of the boundary of~$\Omega$.
 We write $H$ for the corresponding positive self-adjoint operator:
 $H = - \nabla^2$ with boundary conditions understood.
 The eigenvalues of $H$ are positive, with the possible exception 
(in the totally Neumann case) of a constant eigenfunction with 
eigenvalue zero.
 The formulas in this subsection are presented for arbitrary 
spatial dimension~$d$, but in the next section we specialize to 
$d=2$.

The field equation (\ref{fieldeq}) is obtained canonically from the 
curved-space action and Lagrangian 
\begin{equation} S= \int_\Omega L\, \sqrt{|g|} \,d^{d+1}x, 
\qquad
L={\textstyle\frac12}\left[g^{\mu\nu}\,\partial_\mu\phi\,\partial_\nu\phi
  + \xi R\,\phi^2\right],
\label{lagrangian} \end{equation}                      
by taking the variation with respect to $\phi$ and then setting 
the metric to its flat-space value, $g_{\mu\nu}=\eta_{\mu\nu}\,$.
(Our tensorial sign conventions are that $\eta_{00}<0$,
  but $T_{00}>0$ for normal matter.)
 The stress tensor  is defined by
\begin{equation}  T^{\mu\nu} = \frac2 {\sqrt{|g|}} \,
  \frac{\delta S}{\delta g_{\mu\nu}}\,. 
  \label{stressdef}\end{equation}
 It reduces in flat space-time (after  use of the equation of motion,
(\ref{fieldeq})) to
 \begin{equation}
 T_{\mu\nu} = (1-2\xi)\, \partial_\mu\phi\, \partial_\nu\phi
 +\bigl(2\xi-{\textstyle\frac12}\bigr) \eta_{\mu\nu}  \,
\partial_\lambda\phi \,
 \partial^\lambda \phi -2\xi\, \phi\,\partial_\mu\partial_\nu\phi.
 \label{stresst}\end{equation}

  In (\ref{lagrangian}) $R$ is the curvature scalar, and
 $ \xi$
  labels different possible gravitational couplings.
 In curved space different values of $\xi$ are different theories; 
  after the reduction to flat space 
 the field equation is 
independent of~$\xi$, but the stress tensors are different.
 It turns out [see (\ref{T00})] that changing $\xi$ changes  
 $T_{00}$ only 
by a  divergence, and therefore the total  energy
 $  E = \int_\Omega T_{00}\, d\mathbf{r}$
is independent of $\xi$, at least classically,
 under Dirichlet or Neumann boundary conditions.
(A Robin boundary condition \cite{leb,systemat,sah}, 
 $\hat\mathbf{n}\cdot\nabla\phi = \gamma \phi$,
would require a boundary term to be added to the action 
(\ref{lagrangian}).
There results a $\xi$-dependent boundary term in $E$, which 
vanishes when $\xi=\frac14\,$.
Similar remarks apply to models with delta function potentials 
\cite{delta,ines1,ines2}.)
There are three natural choices of~$\xi$:
 \begin{description}
 \item{$\xi=0\,$:} minimal coupling, which simplifies the 
Lagrangian and curved-space field equation;  
\smallskip \item{$\xi = \xi_d\,$:} conformal coupling, 
 \begin{equation}
 \xi_d \equiv \frac{d-1}{4d}\,;
 \qquad
 \xi_2 ={\textstyle \frac18}\,, \quad \xi_3 = {\textstyle\frac16}\,, \quad
 \xi_\infty={\textstyle\frac14}\,,
 \label{conformal}\end{equation}
 which  results in the mildest behavior of the quantized field near 
the boundary; 
\smallskip \item{$\xi = \frac14\,$:} the coupling that
  eliminates the Robin boundary energy, which also
 simplifies the relation between the stress tensor and the total 
energy, as we shall see.
\end{description}

 It is convenient to adopt $\xi=\frac14$ as the base value and to 
define 
 $ \beta= \xi-\frac14 $
 to parametrize the coupling.
 Thus we write
  \begin{equation}
 T_{\mu\nu}(\xi) \equiv T_{\mu\nu}\q + \Delta T_{\mu\nu}
  \label{Tgen}\end{equation}
 and obtain
\begin{equation}
  T_{00}\q =
 \frac12\Biggl[\left(\pd \phi {x^0}\right)^2
 -\phi\nabla^2\phi \Biggr],  \qquad
 \Delta T_{00} = -2\beta\nabla\cdot(\phi\nabla\phi),    
\label{T00}\end{equation}                       
 \begin{equation}
 T_{0j}\q =  \frac12\left[\pd{\phi}{x^0}\,\pd{\phi}{x_j}\,
 -\phi\,\pd{^2\phi}{x^0\,\partial x_j}\right], \qquad
 \Delta T_{0j} = -2\beta\pd{}{x^0}\left(\phi\,\pd{\phi}{x_j}\right),
 \label{T0j}\end{equation}
\begin{eqnarray}
 T_{jk}\q &=& \frac12\left[\pd{\phi}{x_j}\,\pd{\phi}{x_k}
 - \phi\, \pd{^2\phi}{x_j\,\partial x_k} \right], 
 \nonumber\\
 \Delta T_{jk} &=&-2\beta \left[ \pd {\phi}{x_j}\, \pd{\phi}{x_k}
+ \phi\, \pd{^2\phi}{x_j\,\partial x_k} \right]
 \quad\hbox{when $j\ne k$,}
 \label{Tjk}\end{eqnarray}
 \begin{eqnarray}
 T_{jj}\q &=& \frac12\Biggl[\left(\pd{\phi}{x_j}\right)^2 
 - \phi\,\pd{^2\phi}{x_j{}\!^2}\Biggr], 
 \nonumber \\
 \Delta T_{jj}&=& -2\beta\Biggl[\left(\pd{\phi}{x^0}\right)^2 
 -\sum_{k\ne j} \left(\pd{\phi}{x_k}\right)^2
 +\phi\,\pd{^2\phi}{x_j{}\!^2}\Biggr].
 \label{Tjj}\end{eqnarray}
The trace of the tensor is
 \begin{equation}
T^\lambda_\lambda = -\left({\textstyle\frac12}+2\beta d\right)
 \Biggl[\left(\pd{\phi}{x^0}\right)^2 - (\nabla\phi)^2\Biggr],
 \label{trace}\end{equation}
 which vanishes for the conformal coupling, $\beta=- (4d)^{-1}$.

When the theory is canonically quantized, the vacuum expectation 
value  of the stress tensor is expressed formally in terms of the 
 normal modes 
 \[  \varphi_n =\frac1{\sqrt{2\omega_n}} \,\phi_n(\mathbf{r}) 
 e^{-i\omega_n x^0}, \qquad H\phi_n = \omega_n{}\!^2\phi_n\,, 
 \qquad \|\phi_n\|=1,\]
 as
 \begin{equation}
 \langle T_{\mu\nu}(\mathbf{r})\rangle = \sum_{n=1}^\infty
T_{\mu\nu} [\varphi_n,\varphi_n^*].
 \label{modesum}\end{equation}
 (The Neumann zero mode, when it exists, is  omitted and  
ignored. If included and treated properly, it would add a continuous 
energy spectrum \cite[Appendix]{aspects}.)
 The notation in (\ref{modesum}) means that  in each 
of the bilinear terms in (\ref{T00})--(\ref{Tjj}), the field $\phi$
 is replaced by a mode function in one factor and by its 
complex conjugate in the other. (When the factors are not the same, 
the product should be symmetrized.)
 In particular,
 \begin{equation}
 \langle T_{00}\q\rangle  = 
 {\textstyle\frac12}\sum_n \omega_n |\phi_n(\mathbf{r})|^2.
 \label{T00sum}\end{equation}
 Integrating $T_{00}\q$ over $\Omega$ gives the expected formal sum 
for the total energy,
 \begin{equation}
 \langle E\rangle  = {\textstyle\frac12}\sum_n \omega_n \,.
 \label{energysum}\end{equation}

 As promised earlier, we regularize all these divergent sums with 
an exponential ultraviolet cutoff.
 It is convenient to start from the (Poisson) cylinder kernel,
   \begin{equation}
   T(t,\mathbf{r},\mathbf{r}')\equiv 
\sum_{n=1}^\infty  \phi_n(\mathbf{r}) 
 \phi_n(\mathbf{r}')^* e^{-t\omega_n} =
 \langle \mathbf{r}|e^{-t\sqrt{H}} |\mathbf{r}' \rangle.
  \label{cyl}\end{equation}
 (Here $t$ is not the physical time.) Then
 \begin{equation}
  \langle T_{00}\q  \rangle_t 
 = -\, \frac12\,\pd Tt(t,\mathbf{r},\mathbf{r}),
 \label{Toot}\end{equation}
\begin{equation}
 \langle E\rangle_t = -\, \frac12\, \pd{}t 
 T(t), \qquad
  T(t) \equiv \int_\Omega 
 T(t,\mathbf{r},\mathbf{r})\,d\mathbf{r}.
 \label{ET}\end{equation}
To obtain $\langle \Delta T_{00}\rangle $ and the other components of 
$\langle T_{\mu\nu}\rangle $ one needs a more primitive cylinder kernel,
\begin{equation}
 \overline T(t,\mathbf{r},\mathbf{r}') =
- \sum_{n=1}^\infty \frac{1}{\omega_n}\, 
\phi_n(\mathbf{r})\phi_n(\mathbf{r}')^* e^{-t\omega_n}.
 \label{Tbar}\end{equation}
Then $T = \pd {\overline T}t$ and
  \begin{equation}
 \langle \Delta T_{00}\rangle _t = \beta
      \nabla_\mathbf{r}\cdot[\nabla_{\mathbf{r}'}
 \overline T(t,\mathbf{r},\mathbf{r}')]_{\mathbf{r}'=\mathbf{r}}\,.
 \label{DeltaTt}\end{equation}
 In terms of partial differential equations,
 $T$ and $\overline T$ are characterized by the elliptic equation 
 \begin{equation}
\pd{^2T}{t^2} = - \nabla^2 T
 \label{cyleq}\end{equation}
  along with the imposed spatial 
boundary conditions, the initial condition
\[ T(0,\mathbf{r}, \mathbf{r}') = \delta(\mathbf{r}-\mathbf{r}') =
 \pd{\overline T}t(0,\mathbf{r}, \mathbf{r}') ,\]
 and the requirement of boundedness as $t\to+\infty$.
(The Green function $\overline T$ can be introduced 
differently,
either as twice the Euclidean Green function in 
$\mathbf{R}\times \Omega$  with its source on $t=0$,
or through an analytic continuation to imaginary time 
of the Wightman or Feynman two-point function.)

The vacuum expectation values of the summands in (\ref{T0j}) are 
identically zero, as expected from the mode-by-mode time-reversal 
invariance. 
 For the other components one obtains
 \begin{equation}
 \langle T_{jj}\q\rangle_t =
 \frac18\Biggl[ -2\,\pd{^2}{x_j\,\partial x_j'}
+ \pd{^2}{x_j{}\!^2} 
 + \pd{^2}{x'_j{}\!^2} \Biggr]\overline T ,
 \label{TjjT}\end{equation}
 \begin{equation}
 \langle \Delta T_{jj} \rangle_t =
 \frac\beta2 \Biggl[2\,\pd{^2}{t^2}
  -2\sum_{k\ne j} \pd{^2}{x_k\,\partial x'_k} + \pd{^2}{x_j{}\!^2} 
 + \pd{^2}{x'_j{}\!^2} \Biggr] \overline T ,
 \label{DTjjT} \end{equation}
 \begin{equation}
\langle T_{jk}\q\rangle_t =\frac18\Biggl[
\pd{^2}{x_j\,\partial x_k} +\pd{^2}{x'_j\,\partial x'_k}
 -\pd{^2}{x_j\,\partial x'_k} -\pd{^2}{x'_j\,\partial x_k} 
\Biggr]\overline T,
 \label{TjkT}\end{equation}
 \begin{equation}
\langle \Delta T_{jk}\rangle_t =\frac{\beta}2 \Biggl[
\pd{^2}{x_j\,\partial x_k} +\pd{^2}{x'_j\,\partial x'_k}
 +\pd{^2}{x_j\,\partial x'_k} +\pd{^2}{x'_j\,\partial x_k} 
\Biggr]\overline T,
 \label{DTjkT}\end{equation}
 where it is understood that $\mathbf{r'}$ is to be set equal to 
$\mathbf{r}$ at the final step.

\section{The stress tensor}
\label{sec:local}

 \subsection{Preliminaries}\label{ssec:prelim}

 We now restrict attention to dimension~$2$ and write $x$ for 
$x_1$ and $y$ for $x_2\,$.
 Define
 \begin{eqnarray}
A&\equiv& \pd Tt = \pd{^2\tbar}{t^2}\,, \label{Adef}\\
 B_1 &\equiv& \frac12 \left( \pd{^2\tbar}{x^2} + \pd{^2\tbar}{x'^2} 
\right), \quad 
 B_2 \equiv \frac12 \left( \pd{^2\tbar}{y^2} + \pd{^2\tbar}{y'^2} 
\right), \label{Bdef} \\
 C_1&\equiv& \pd{^2\tbar}{x\,\partial x'}\,,\quad 
C_2 \equiv \pd{^2\tbar}{y\,\partial y'}\,,\quad  \label{Cdef} \\
D_{12}&\equiv& \frac12\left(\pd{^2\tbar}{x\,\partial y'} +
 \pd{^2\tbar}{y\,\partial x'}\right), \label{Ddef} \\
 E_{12} &\equiv& \frac12\left( \pd{^2\tbar}{x\,\partial y} +
\pd{^2\tbar}{x'\,\partial y'}\right)\,. \label{Edef}
 \end{eqnarray}
 (The subscripts on $D$ and $E$ are merely to facilitate later 
generalization to higher dimensions.)
 Then from (\ref{Toot}),  (\ref{DeltaTt}), and 
 (\ref{TjjT})--(\ref{DTjkT}) we have
 \begin{eqnarray} 
\bra T_{00}\q\ket_t &=& -{\textstyle\frac12}A, \label{00calc} \\
\bra \Delta T_{00}\ket_t &=& \beta (B_1 +B_2 +C_1 + C_2), 
 \label{d00calc}\\
 \bra T_{01}\q\ket_t &=& 0 =  \bra \Delta T_{01}\ket_t \,, 
 \quad\hbox{etc.},\label{01calc} \\
 \bra T_{11}\q\ket_t &=& {\textstyle\frac14} (B_1 - C_1), \quad\hbox{etc.},
 \label{11calc} \\
\bra \Delta T_{11}\ket_t &=& \beta (A + B_1 - C_2),  \quad\hbox{etc.}, 
 \label{d11calc}\\
\bra T_{12}\q\ket_t &=& {\textstyle\frac14}(-D_{12}+E_{12}),
 \label{12calc}\\
 \bra \Delta T_{12}\ket_t &=& \beta (D_{12}+E_{12}). \label{d12calc}
\end{eqnarray}
 In (\ref{00calc})--(\ref{d12calc}) it is understood that 
$\mathbf{r}'=\mathbf{r}$.

 \goodbreak

 \subsection{Path classes and energy density} \label{ssec:paths}

 The  cylinder kernels in infinite two-dimensional space are
 \begin{equation}
\tbar(t,\mathbf{r},\mathbf{r}') = -\, \frac1{2\pi}\,
 (t^2 +|\mathbf{r} -\mathbf{r}'|^2)^{-1/2},
 \label{freecylbar}\end{equation}
 \begin{equation} T(t,\mathbf{r},\mathbf{r}') =
 \frac t{2\pi}\, (t^2 +|\mathbf{r} -\mathbf{r}'|^2)^{-3/2}.
 \label{freecyl} \end{equation}

\begin{figure}
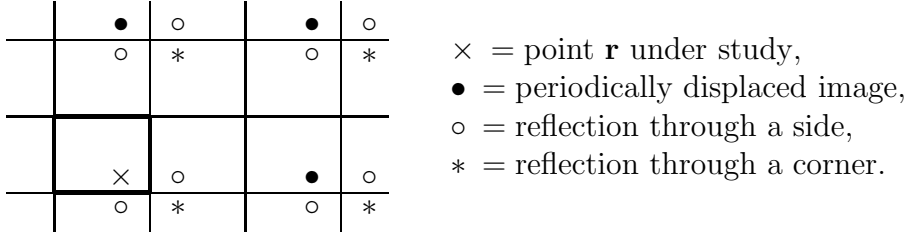

\centerline{\beginpicture
  \setcoordinatesystem units <.5in, .4in> point at 0 1
 \setplotarea x from -0.5 to 3.5, y from  -0.5 to 2.5
 \putrule from -0.5 0 to 3.5 0
  \putrule from -0.5 1 to 3.5 1
 \putrule from -0.5 2 to 3.5 2
 \putrule from 0 -0.5  to 0 2.5
 \putrule from 1 -0.5  to 1 2.5 
 \putrule from 2 -0.5  to 2 2.5 
 \putrule from 3 -0.5  to 3 2.5 
 \put{$\times$} at .7 .2                               
 \put{$\bullet$} at 2.7 .2 
 \put{$\bullet$} at .7 2.2 
  \put{$\bullet$} at 2.7 2.2 
 \put{$\circ$} at 1.3 .2 
  \put{$\circ$} at 3.3 .2
   \put{$\circ$} at .7 -.2
  \put{$\circ$} at 2.7 -.2
 \put{$\circ$} at 1.3 2.2 
  \put{$\circ$} at 3.3 2.2
   \put{$\circ$} at .7 1.8
  \put{$\circ$} at 2.7 1.8
 \put{$*$} at 1.3 -.2 
  \put{$*$} at 3.3 -.2 
 \put{$*$} at 1.3 1.8 
 \put{$*$} at 3.3 1.8 
  \linethickness=1.3pt
 \putrule from 0 0 to 1 0  
  \putrule from 0 1 to 1 1  
 \putrule from 0 0 to 0 1   
 \putrule from 1 0 to  1 1 
 \endpicture\qquad\parbox{2.5truein}
{\begin{description}
 \item[$\times$] $ = \hbox{point $\mathbf{r}$ under study}$,
 \item[$\bullet$] $ = \hbox{periodically displaced image}$,
\item[$\circ$] $ = \hbox{reflection through a side}$,
\item[$*$] $ = \hbox{reflection through a corner}$. 
\end{description}}}
 \caption{A point in a rectangle and its images relevant to 
Dirichlet and Neumann boundary conditions
 (cf.\ \cite[Sec.~9.A]{BB3}). 
 Image points fall into three classes according to whether the 
number of reflections is \emph{even} in both dimensions, one, or 
neither.
 The first case corresponds to periodic displacements.
Points of  the third class are joined to $\mathbf{r}$ by lines that 
pass through an intersection point of the lattice of extended
rectangle sides --- i.e., an image of a corner of the rectangle.}          
\label{fig:images} \end{figure}

 Because of its central importance, we shall discuss the energy 
density, $\bra T_{00}\q\ket$, along with the construction of the 
cylinder kernel for the rectangle, path by path.
 For rectangular parallelepipeds of any dimension,
 with any combination of Dirichlet, Neumann, and periodic boundary 
conditions,
 the construction of any kernel (Green function) as a sum over 
classical paths reduces to the classic ``method of images'' 
 (and yields the exact answer).
For a rectangle the array of image points appears in 
Fig.~\ref{fig:images}.
To every path is associated a sign, $(-1)^\eta$, where $\eta$ is 
the number of Dirichlet sides struck by the path.  
 (If a path hits a corner, both sides are counted, and the path 
reflects back upon itself.)
The image sum for $\overline T$ is not absolutely convergent, but 
the
derivatives of the series, from which observable quantities are
calculated, are convergent.

\begin{figure}
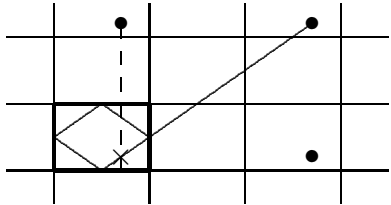
              
 \centerline{\beginpicture
  \setcoordinatesystem units <.5in, .35in> point at 0 0
 \setplotarea x from -0.5 to 3.5, y from  -0.5 to 2.5
 \putrule from -0.5 0 to 3.5 0
  \putrule from -0.5 1 to 3.5 1
 \putrule from -0.5 2 to 3.5 2
 \putrule from 0 -0.5  to 0 2.5
 \putrule from 1 -0.5  to 1 2.5 
 \putrule from 2 -0.5  to 2 2.5 
 \putrule from 3 -0.5  to 3 2.5 
 \put{$\times$} at .7 .2                               
 \put{$\bullet$} at 2.7 .2 
 \put{$\bullet$} at .7 2.2 
  \put{$\bullet$} at 2.7 2.2 
 \plot .7 .2 
 2.7 2.2 /
 \plot 1 .5
 .5  1 /     
 \plot .5 1
 0 .5 /
 \plot 0 .5
  .5 0 /
 \plot .5 0
  .7 .2 /
 \setdashes
 \putrule from .7 0 to .7 2.2 
 \setsolid \linethickness=1.3pt \noindent
 \putrule from 0 0 to 1 0  
  \putrule from 0 1 to 1 1  
 \putrule from 0 0 to 0 1   
 \putrule from 1 0 to  1 1 
 \endpicture}
 \caption{Two periodic paths (one solid, one dashed) are shown, 
both within the rectangle and in the covering space.}
 \label{fig:per}\end{figure}

 Following Cavalcanti \cite{Cav} we take the rectangle to have 
horizontal and vertical dimensions $a$ and $b$, horizontal and 
vertical coordinates $x$ and $y$, and horizontal image-displacement 
indices $j$ and~$k$. 
 (We occasionally still  find it necessary to use $j$ and $k$ as 
tensor indices, but never in the same equation as the image 
indices.)
 Thus the contribution of a typical periodic 
path (see Fig.~\ref{fig:per}) to $\tbar$ is 
 \begin{equation}
\tbar_{\mathrm{P}jk} = -\,\frac{(-1)^\eta}{2\pi} 
[t^2 + (2ja +x'-x)^2 + (2kb+y'-y)^2)]^{-1/2}.
\label{Pjk}\end{equation}
From (\ref{00calc}) and (\ref{Adef}) we obtain
\begin{eqnarray}
\bra T_{00}\q\ket_{t\mathrm{P}jk} &=&
  -\,\frac{(-1)^\eta}{4\pi} \left[t^2+(2ja)^2 +(2kb)^2\right]^{-5/2}
 \nonumber\\ &&\times
 \left[-2t^2+(2ja)^2 + (2kb)^2\right],
 \label{T00Pjk}\end{eqnarray}
 which is independent of~$\mathbf{r}$.
  Also, one finds from (\ref{d00calc}) and 
 (\ref{Bdef})--(\ref{Cdef})
 that $C_j = - B_j$ in this case and hence
 \begin{equation}
 \bra \Delta T_{00}\ket_{t\mathrm{P}jk} = 0.
 \label{dT00Pjk}\end{equation}
 These two results are expected and related:
 Since $\Delta T_{00}$ is a total divergence and hence must 
integrate to $0$, and since the energy from  a periodic path is 
independent of position in the rectangle, 
 $\bra \Delta T_{00}\ket_{t\mathrm{P}jk}$ must be identically zero.
Finally, note that  if the boundaries are all Dirichlet or all Neumann, 
 $\eta$ is even and hence $\bra T_{00}\ket_{0\mathrm{P}jk}$ is 
always negative.

  \begin{figure}
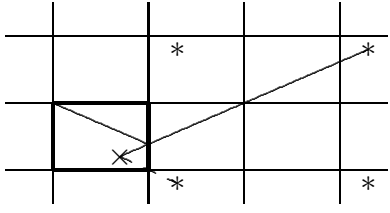

  \centerline{\beginpicture
  \setcoordinatesystem units <.5in, .35in> point at 0 0
 \setplotarea x from -0.5 to 3.5, y from  -0.5 to 2.5
 \putrule from -0.5 0 to 3.5 0
  \putrule from -0.5 1 to 3.5 1
 \putrule from -0.5 2 to 3.5 2
 \putrule from 0 -0.5  to 0 2.5
 \putrule from 1 -0.5  to 1 2.5 
 \putrule from 2 -0.5  to 2 2.5 
 \putrule from 3 -0.5  to 3 2.5 
 \put{$\times$} at .7 .2                               
 \put{$*$} at 1.3 -.2 
  \put{$*$} at 3.3 -.2 
 \put{$*$} at 1.3 1.8 
 \put{$*$} at 3.3 1.8 
 \plot .7 .2
 3.3 1.8 /
 \plot  1 .3846
        0 1 /
  \setdashes
 \plot .7 .2
    1.3 -.2 /
 \setsolid \linethickness=1.3pt \noindent
 \putrule from 0 0 to 1 0  
  \putrule from 0 1 to 1 1  
 \putrule from 0 0 to 0 1   
 \putrule from 1 0 to  1 1 
 \endpicture}
 \caption{Two corner paths.  Inside the rectangle such paths 
bounce back from a corner and retrace themselves.
 The shortest such paths have lengths arbitrarily close to $0$.}
\label{fig:corner} \end{figure}

Next, consider the simple and interesting case of a corner path:
 \begin{equation}
 \tbar_{\mathrm{C}jk}
 = -\,\frac{(-1)^\eta}{2\pi} 
 \left[t^2 +(2ja -x'-x)^2 +(2kb -y'-y)^2  \right]^{-1/2}.
 \label{Cjk}\end{equation}
 This time one finds that $C_j = +B_j$ and $B_1+B_2 =-A$, so that
\begin{eqnarray}
  \bra T_{00}\q\ket_{t\mathrm{C}jk} &=&
{\textstyle \frac1{4\beta}} \bra\Delta T_{00}\ket_{t\mathrm{C}jk}
 \nonumber \\  &=&
 -\, \frac{(-1)^\eta}{4\pi}\left[t^2+(2ja-2x)^2 +(2kb-2y)^2\right]^{-5/2}
 \nonumber \\ &&{}\times 
 \left[-2t^2+(2ja-2x)^2 + (2kb-2y)^2\right] .
 \label{T00Cjk}\end{eqnarray}
 That is, the two terms in $\bra T_{00}\ket_{t\mathrm{C}jk}$ are 
proportional, and, in particular,\break   
 $\bra T_{00}\ket_{t\mathrm{C}jk}$ vanishes 
for minimal coupling ($\beta=-\frac14$).
 These seeming coincidences are probably related to the fact that 
 the integral of $\bra T_{00}\ket_{tCjk}$ over the rectangle must 
vanish (see Sec.~\ref{sec:global}).  Note that the quantity is a 
function  of the distance to $\mathbf{r}$  from the corner or 
corner-image concerned (see Fig.~\ref{fig:corner}).
 Again it is negative as $t\to0$
 whenever all the boundary conditions are of 
the same type.
                                       
    \begin{figure}
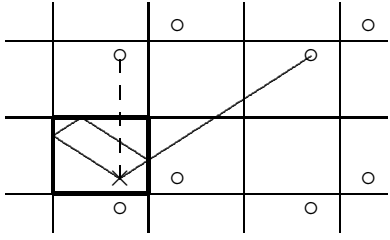

 \centerline{\beginpicture
  \setcoordinatesystem units <.5in, .4in> point at 0 0
 \setplotarea x from -0.5 to 3.5, y from  -0.5 to 2.5
 \putrule from -0.5 0 to 3.5 0
  \putrule from -0.5 1 to 3.5 1
 \putrule from -0.5 2 to 3.5 2
 \putrule from 0 -0.5  to 0 2.5
 \putrule from 1 -0.5  to 1 2.5 
 \putrule from 2 -0.5  to 2 2.5 
 \putrule from 3 -0.5  to 3 2.5 
 \put{$\times$} at .7 .2                               
 \put{$\circ$} at 1.3 .2 
  \put{$\circ$} at 3.3 .2
   \put{$\circ$} at .7 -.2
  \put{$\circ$} at 2.7 -.2
 \put{$\circ$} at 1.3 2.2 
  \put{$\circ$} at 3.3 2.2
   \put{$\circ$} at .7 1.8
  \put{$\circ$} at 2.7 1.8
 \plot
 0.7 0.2
 2.7 1.8 /
 \plot 1 0.44
 0.3 1 /
 \plot 0.3 1
 0 0.76 /
 \plot 0 0.76 
 0.7 0.2 /
 \setdashes
 \putrule from 0.7 0.2 to 0.7 1.8
 \setsolid \linethickness=1.3pt \noindent
 \putrule from 0 0 to 1 0  
  \putrule from 0 1 to 1 1  
 \putrule from 0 0 to 0 1   
 \putrule from 1 0 to  1 1 
 \endpicture}
 \caption{Two side paths of the vertical subclass.
   The dashed path is a direct 
perpendicular reflection (retracing itself), 
 with length approaching $0$ as 
$\mathbf{r}$ approaches the top boundary.
 The solid path combines a reflection from the top with a periodic 
horizontal drift; its length is bounded away from $0$.}
\label{fig:side} \end{figure}

 The situation is slightly more complicated for paths that 
 ``bounce'' in one dimension while being periodic (or fixed)
 in the other.
 The number of reflections is now odd, and the energy density
  turns out to be 
positive for Dirichlet conditions and negative for Neumann,
 at least for $\beta$ near~$0$.
 We call these paths ``vertical side paths'' if
 the bounce is off a horizontal side (see Fig.~\ref{fig:side});
 this includes, in particular, the strictly vertical paths
($j=0$).
 In those cases we have
 \begin{equation}
   \tbar_{\mathrm{V}jk} = -\,\frac{(-1)^\eta}{2\pi} 
 \left[t^2 +(2ja +x'-x)^2 +(2kb - y'-y)^2  \right]^{-1/2},
 \label{Vjk}\end{equation}                                      
 \begin{eqnarray}
   \bra T_{00}\q\ket_{t\mathrm{V}jk}    & =& 
-\,  \frac{(-1)^\eta}{4\pi}\left[t^2+(2ja)^2 +(2kb-2y)^2\right]^{-5/2} 
  \nonumber\\&& {}\times
 \left[-2t^2+(2ja)^2 + (2kb-2y)^2\right] ,
 \label{T00Vjk}\end{eqnarray}
 \begin{eqnarray}
 \bra \Delta T_{00}\ket_{t\mathrm{V}jk}   & =& 
 \frac{\beta(-1)^\eta}{\pi} 
  \left[t^2+(2ja)^2 +(2kb-2y)^2\right]^{-5/2} 
\nonumber \\ && {}\times
 \left[t^2+  (2ja)^2 -2 (2kb-2y)^2\right]. 
 \label{dT00Vjk}\end{eqnarray}                 
  These quantities depend only on $y$, not~$x$;
 in other words, such a term is a function of the 
distance from a wall or an image of a wall.           
 In this case the two terms in the energy density are distinctly 
different, so it pays to write out the total explicitly:
 \begin{eqnarray}
   \bra T_{00}\ket_{t\mathrm{V}jk}    & = &
 \frac{(-1)^\eta}{\pi}  \left[t^2+(2ja)^2 +(2kb-2y)^2\right]^{-5/2} 
\nonumber\\&& {}\times
 \left[\left(\beta +{\textstyle\frac12}\right)t^2 
 + \left(\beta-{\textstyle\frac14}\right)(2ja)^2
  - \left(2\beta+{\textstyle\frac14}\right)(2kb-2y)^2\right].
 \label{totT00Vjk}\end{eqnarray}
 The most interesting observation here is that
the coefficient of $(2kb-2y)^2$  vanishes
for conformal coupling ($\beta=-\frac18$).
 When $j=0$ and $k=0$ or $1$, the energy density for $t=0$ 
 generically has 
  $O\left(y^{-3}\right)$ divergences at the boundary, but those 
  divergences are removed in the conformal case;
 this is as close as one comes in a rectangle to the well known 
fact that the energy density between infinite parallel plates is 
\emph{constant} in the case of conformal coupling.

  Formulas for horizontal side paths are easily obtained from 
 (\ref{Vjk})--(\ref{totT00Vjk}) by 
interchanging the roles of the two dimensions.

\subsection{The other components} \label{ssec:Tjk}

From (\ref{11calc})--(\ref{d12calc})
 and (\ref{Pjk}), (\ref{Cjk}), (\ref{Vjk}),
  one finds the spatial 
components (pressure and shear stress).  We omit the formula for 
the 22 component when it is obvious from the 11 formula. 

\emph{Periodic paths:}
\begin{eqnarray}
   \langle T_{11}\q\rangle_{t\mathrm{P}jk} &=& 
\frac{(-1)^\eta}{4\pi} 
 \left[t^2+(2ja)^2 +(2kb)^2\right]^{-5/2} 
 \nonumber \\
  &&{}\times\left[t^2-2(2ja)^2 +(2kb)^2\right] ,
 \label{T11Pjk}\end{eqnarray}
\begin{equation}
   \langle T_{12}\q\rangle_{t\mathrm{P}jk} = 
-\,\frac{3(-1)^\eta}{\pi} 
 \left[t^2+(2ja)^2 +(2kb)^2\right]^{-5/2} 
 jakb,
 \label{T12Pjk}\end{equation}
 \begin{equation}
  \langle \Delta T_{11}\rangle_{t\mathrm{P}jk} =0 =
\langle\Delta T_{12}\rangle_{t\mathrm{P}jk}\,.
 \label{dT11Pjk}\end{equation}
 Thus the stress tensor associated with a periodic path does not 
depend upon the conformal parameter, nor upon the coordinates.  The 
individual terms $\langle T_{12}\rangle$  are nonzero, but they add 
to zero when summed over either $j$ or~$k$, as reflection symmetry 
requires.

\emph{Corner paths:}
 \begin{equation}
  \langle T_{11}\q\rangle_{t\mathrm{C}jk}
 = \langle T_{12}\q\rangle_{t\mathrm{C}jk}=0 ,
 \label{T11Cjk}\end{equation}
 \begin{eqnarray}
 \langle \Delta T_{11}\rangle_{t\mathrm{C}jk} &=& 
  -\,\frac{\beta(-1)^\eta}{\pi} 
 \left[t^2+(2ja-2x)^2 +(2kb-2y)^2\right]^{-5/2} \nonumber \\
  &&{}\times
 \left[t^2+(2ja-2x)^2 -2 (2kb-2y)^2\right] ,
 \label{dT11Cjk} \end{eqnarray}
 \begin{eqnarray}
   \langle \Delta T_{12}\rangle_{t\mathrm{C}jk} &=&
 -\,\frac{12\beta(-1)^\eta}{\pi}
   \left[t^2+(2ja-2x)^2 +(2kb-2y)^2\right]^{-5/2} 
 \nonumber\\ &&\times
 (ja-x)(kb-y).
\label{dT12Cjk} \end{eqnarray} 
 In addition to (and in contrast to) the remarks about the energy 
density made below (\ref{T00Cjk}), 
 we observe:
 (1) The spatial components of the corner-path stress tensor vanish 
when $\xi=\frac14$ (whereas the energy density vanishes when 
$\xi=0$). So far we have no intuitive explanation of this fact.
 (2) The spatial components are no longer functions of corner-image 
distances alone, though they do have (for $t=0$) an  
$O(|\mathbf{r}|^{-3})$ dependence on 
 corner-image coordinates, as the energy density does.
 (3)       When $\beta\ne0$ there is a nonzero 
 $\langle T_{12}\rangle$, which does 
not vanish even when summed.  However, if we evaluate it on a 
boundary (such as $x =\hbox{(integer)} \times a$), where it would 
have a clear physical interpretation as a shear force on the wall 
of the box, then it does vanish when summed.

\emph{Vertical paths:}
 \begin{eqnarray}
   \langle T_{11}\q\rangle_{t\mathrm{V}jk} &=& 
\frac{(-1)^\eta}{4\pi} 
  \left[t^2+(2ja)^2 +(2kb-2y)^2\right]^{-5/2} 
 \nonumber\\ &&\times
 \left[t^2-2(2ja)^2 +(2kb-2y)^2\right] ,
 \label{T11Vjk}\end{eqnarray}
\begin{eqnarray}
    \langle \Delta T_{11}\rangle_{t\mathrm{V}jk} &=& 
-\,\frac{\beta(-1)^\eta}{\pi}\left[t^2+(2ja)^2 +(2kb-2y)^2\right]^{-5/2} 
 \nonumber\\  &&{}\times
  \left[t^2 +(2ja)^2-2(2kb-2y)^2 \right],
\label{dT11Vjk} \end{eqnarray}
 \begin{equation} 
 \langle T_{22}\rangle_{t\mathrm{V}jk} = 0  = 
 \langle T_{12}\rangle_{t\mathrm{V}jk} = 0.
 \label{T22Vjk} \end{equation}
 In addition to the remarks surrounding 
(\ref{Vjk})--(\ref{totT00Vjk}), 
observe that   $\langle T_{\nu2}\rangle = 0$ for all~$\nu$.
  That is understandable:  there is otherwise no way to 
satisfy the conservation laws (\ref{conslaw}) for 
 $\mu=1$ and $\mu=2$ 
by functions that depend only on $y$ but are not constant.

\emph{Horizontal paths:}
 \begin{equation} 
 \langle T_{11}\rangle_{t\mathrm{H}jk} = 0 
  = \langle T_{12}\rangle_{t\mathrm{H}jk} = 0,
 \label{T12Hjk} \end{equation}
 \begin{eqnarray}
   \langle T_{22}\q\rangle_{t\mathrm{H}jk} &=& 
\frac{(-1)^\eta}{4\pi} 
  \left[t^2+(2ja-2x)^2 +(2kb)^2\right]^{-5/2} 
 \nonumber\\ &&{}\times
 \left[t^2+(2ja-2x)^2 -2(2kb)^2\right] ,
 \label{T22Hjk}\end{eqnarray}
\begin{eqnarray}
    \langle \Delta T_{22}\rangle_{t\mathrm{H}jk} &=& 
-\,\frac{\beta(-1)^\eta}{\pi}  \left[t^2+(2ja-2x)^2 +(2kb)^2\right]^{-5/2} 
 \nonumber\\ &&{}\times
 \left[t^2 -2(2ja-2x)^2+(2kb)^2 \right].
\label{dT22Hjk} \end{eqnarray}
 Observe that $\langle T_{12} \rangle=0$ for \emph{all} side 
paths.

For the formulas above one can verify the conservation law
 \begin{equation}
 -\,\pd{}{x^0} \langle T_{0\mu}\rangle +\pd{}{x_1} \langle 
T_{1\mu}\rangle
+ \pd{}{x_2} \langle T_{2\mu}\rangle =0 \quad (\mu=0,1,2).
 \label{conslaw}\end{equation}
 Here the first term is always $0$, because the quantities do not 
depend upon time (not to be confused with the regularization 
parameter~$t$).
 In the conformal case, $\beta=-\frac18\,$, one also has 
 tracelessness,
 \begin{equation}
 -\,\langle T_{00}\rangle + \langle T_{11}\rangle +
  \langle T_{22}\rangle=0.
 \label{tracelaw}\end{equation}
 These identities hold for all $t$, not just $t=0$.

\section{Energy and force}
\label{sec:global}

 \subsection{Introductory remarks}\label{ssec:introrem}

 In this section the results of the previous one will be used to 
calculate the contribution of each image term to the total energy, 
$E$, of the scalar field in the rectangle, and consequently the 
force, $-\pd{E}{a}$, on the rectangle's right side.
 We are concerned here only with the force from inside the 
rectangle; ``piston'' arrangements in which it is possible to 
calculate or estimate forces from outside will be considered in 
later sections.
 Consequently, uncompensated divergent terms  arise as the 
cutoff parameter $t$ is taken to~$0$, and such terms  need to 
be identified and systematically isolated for later physical 
scrutiny.

 The sign of Casimir energies and forces has long been a topic of 
great interest and mystery, and one of the motivations of our 
research has been to see what light the decomposition into image 
terms, for which the sign is easy to understand, can shed on such 
questions.  The following discussion is easy to present for 
arbitrary spatial dimension~$d$. 
 
  The cylinder kernel in $\mathbf{R}^d$
is
 \begin{equation}
T(t,\mathbf{r},\mathbf{r}')= 
 C(d)\, t (t^2+|\mathbf{r}-\mathbf{r}'|^2)^{-(d+1)/2},
 \qquad
 C(d) \equiv \frac {\Gamma(\frac{d+1}2)} {\pi^\frac{d+1}{2}}\,.
\label{cylgendim}\end{equation}
Consequently, in the $d$-dimensional analogues of the constructions 
in Sec.~\ref{ssec:paths} all the terms in the energy density 
 $\bra T_{00}\q\ket$
 will have the form 
 \begin{equation}
-\,\frac{(-1)^\eta}{2}\, \pd{}t [t(t^2+W)^{-s}]
 = (-1)^\eta \,\frac {(s-\frac12)t^2 -\frac12 W}{(t^2+W)^{s+1} } 
\,,
\label{airportterm} \end{equation} 
   where $W$ is some nonnegative function of $\mathbf{r}$.
 If $W>0$, the limit as $t\to0$ is 
 $-\frac12 (-1)^\eta W^{-s}$; for pure Neumann boundary conditions 
it is always negative, while for pure Dirichlet conditions it will 
be positive whenever the number of nonperiodic (bounce) dimensions 
is odd for the path concerned.
 If $W=0$, the small-$t$ behavior is
 $(-1)^\eta (s-\frac12) t^{-2s} $,
 divergent and opposite in sign to the other terms.

 Now we consider integrating over a coordinate $u$ when
$ W = V + (mL-u)^2$ with $V\ge 0$:
 \[ I \equiv \int_0^L \frac
 {(s-\frac12) t^2 -\frac12 [V+(mL-u)^2] } {
 [t^2+V+(mL-u)^2]^{s+1} } \, du. \]
By the mean value theorem for integrals,
 \[I= L \,\frac{(s-\frac12) t^2 -\frac12 [V+(mL-\zeta)^2] } {
 [t^2+V+(mL-\zeta)^2]^{s+1} } \,,\]
 where $0<\zeta<L$ and $\zeta$ may depend on $t$.
 But for us $s$ is always greater than or equal to~$1$.
 So if the integral converges at all, the integrand is bounded and 
we are back to the situation of the previous paragraph with 
 $W(t)>0$ and having a positive lower bound.
 In the contrary case, $V=0$ and $m=0$ or $1$,
 the situation is more delicate but the sign question is absorbed 
into the issue of the physical meaning of  surface divergences.

 In summary, for all \emph{finite} terms we have good control over 
the sign.
 (The  $\bra\Delta T_{00}\ket$ terms are irrelevant to total energy, 
as discussed below.)
 One  understands why Ambj\o{}rn and Wolfram \cite[Table~I]{AW} found 
nontrivial 
signs only for Dirichlet problems (not periodic or Neumann), and the 
particular sign patterns they saw are not surprising.

 To understand the significance of various paths, it is useful to 
refine the classification of paths in the previous section.
 Each closed path is characterized by its image indices, $j$ 
 and~$k$, and by its periodicity type, P, V, H, or C.

 \begin{description}
 \item[P:] Periodic paths, producing constant terms in the energy 
density
\smallskip \begin{description}
 \item[PZ:] $j=0=k$ ---  the zero-length path          
 \item[PV:] $j=0$, $k\ne0$ ---  vertical periodic paths
  \item[PH:] $k=0$, $j\ne0$ --- horizontal periodic paths
 \item[PD:]  $j\ne0$, $k\ne0$ --- diagonal periodic paths
 \end{description} \medskip\goodbreak
 \item[V:] Nonperiodic closed paths whose uncompensated ``bounce'' 
occurs on the top or bottom side of the rectangle, producing energy 
densities depending on $y$ only
 \smallskip\begin{description}
\item[VP:] $j=0$ --- perpendicular vertical bounce paths
 \item[VD:] $j\ne 0$ --- vertical bounce paths with horizontal 
periodic drift
  \end{description}  \medskip\goodbreak
 \item[H:] Nonperiodic closed paths whose uncompensated ``bounce'' 
occurs on the right or left side of the rectangle, producing energy 
densities depending on $x$ only
 \smallskip\begin{description}
\item[HP:] $k=0$ --- perpendicular horizontal bounce paths
 \item[HD:] $k\ne 0$ --- horizontal bounce paths with vertical 
periodic drift
  \end{description} \medskip
 \item[C:] Closed paths that are periodic in neither dimension,
  producing  energy densities associated with corner images
 \end{description}

 Path PZ produces, by (\ref{ET}), the ubiquitous 
volume (here area) divergence, 
 \begin{equation}
T_\mathrm{PZ}(t) = \frac{ab}{2\pi t^2}\,, \qquad
 \langle E\rangle_{t\mathrm{PZ}} = \frac{ab}{2\pi t^3}\,,
 \label{Zterm}\end{equation}
 which, being ubiquitous, is always ignored 
 (except for possible relevance to cosmological dark energy).
All other terms in the energy density are pointwise finite, but 
some of them have nonintegrable divergences at the boundary.
  The path classes involved are VP and HP, which produce the 
well known  surface (here perimeter) divergence in the total 
energy, and C, which produces an  energy density that 
seemingly diverges at the corners but nevertheless makes no 
contribution to the ``renormalized'' total energy, 
  as we shall see.

 \subsection{Energy calculations}

 Let us first dispose of 
 $\int\!\!\int\bra\Delta T_{00}\ket\, dx\,dy$, 
 which is expected to be zero because $\Delta T_{00}$ is the 
divergence of a vector field, proportional to $\phi\nabla\phi$,
  that vanishes on every Dirichlet or Neumann boundary.
 From (\ref{dT00Pjk}) and (\ref{T00Cjk}) we see that the quantity 
is indeed zero for periodic paths, while for corner paths it is 
proportional to the $T_{00}\q$ term  (which also will turn out to be 
zero).
 The situation for side paths is more subtle.
 The integral of $\bra\Delta T_{00}\ket_{t\mathrm{V}jk}$
 from (\ref{dT00Vjk}) is not zero, which is not surprising since 
the field from a single image source does not satisfy the boundary 
conditions.
 However, because (\ref{dT00Vjk}) is a total derivative, 
 a calculation almost identical to that in 
 (\ref{Delcalc1})--(\ref{Delcalc2}) below 
 shows that the sum over $k$ does 
telescope to~$0$, at least when the top and bottom boundaries are 
of the same type (both Dirichlet or both Neumann).

The total energy contributed by a periodic path is trivially 
obtained by multiplying (\ref{T00Pjk}) by the area, $ab$.
 The sum of all such terms splits into PV, PH, and PD parts as
 \begin{eqnarray} \langle 
E\rangle_{t\mathrm{P}\setminus\mathrm{Z}}
&=&-\,\frac{ab}{2\pi} \sum_{k=1}^\infty (-1)^\eta \frac{(2kb)^2 -2t^2}
 {[t^2 +(2kb)^2]^{5/2}}  
-\frac{ab}{2\pi} \sum_{j=1}^\infty (-1)^\eta \frac{(2ja)^2 -2t^2}
 {[t^2 +(2ja)^2]^{5/2}}  
 \nonumber\\
&&{} - \frac{ab}{\pi} \sum_{j=1}^\infty\sum_{k=1}^\infty (-1)^\eta 
 \frac{(2ja)^2 +(2kb)^2 -2t^2}{[t^2 +(2ja)^2 +(2kb)^2]^{5/2}}\,.
 \label{regperenergy}\end{eqnarray}
If all four sides are of the same type, (\ref{regperenergy}) 
simplifies in the limit $t\to0$ to
 \begin{equation}
 \langle E\rangle_{t\mathrm{P}\setminus\mathrm{Z}}=
 -\,\frac{\zeta(3)a}{16\pi b^2} - \frac{\zeta(3)b}{16\pi a^2}
-\,\frac{ab}{8\pi} \sum_{j=1}^\infty\sum_{k=1}^\infty
 \left(a^2j^2 + b^2k^2\right)^{-3/2}
 \label{perenergy}\end{equation}
 (a  well known result --- e.g., \cite{EORBZ}).
 In the special case of a square of side~$a$, 
 numerical evaluation of (\ref{perenergy}) gives $-0.089859/a$
 (identifiable with the vacuum energy of a torus of dimension $2a$ 
as recorded in \cite{AW, EORBZ}).
Because of the need to sum over a two-dimensional lattice,
  the numerical convergence is rather slow, even when repetitions of 
primitive orbits are handled all at once ---
 in contrast with the situation for parallel plates, where the sum 
over paths has been found to be very efficient and increasingly so 
in higher dimensions \cite{JS,LF}.

 To calculate the other $T_{00}\q$ terms it is convenient to return 
to the cylinder kernel $T$ 
 and integrate it over the rectangle before taking the final $t$ 
derivative in~(\ref{Adef}).

 Consider first the paths of subclass VP. 
 According to (\ref{Vjk}) with $\mathbf{r}'=\mathbf{r}$,
 for any path of class~V we have
 \begin{equation}
 T_{\mathrm{V}jk}(t,\mathbf{r},\mathbf{r}) =
  +\,\frac{(-1)^\eta}{2\pi} \,t
 \left[t^2 + (2ja)^2 + (2kb-2y)^2\right]^{-3/2}.
 \label{Vcyl}\end{equation}
 Setting $j=0$ we arrive at
 \[  \int_0^a dx \int_0^b dy
  \,T_{\mathrm{V}0k}(t,\mathbf{r},\mathbf{r}) =
\frac{(-1)^\eta}{2\pi} \,at \int_0^b
 \left[t^2  + (2kb-2y)^2\right]^{-3/2}\,dy. \]
 The terms with $k=0$ and $k=1$ are divergent (when $t\to0$)
 at the bottom and top boundaries, respectively.
 The other cases are finite, but need to be added to the divergent 
ones to build up a ``clean''  divergence, proportional to a power 
of~$t$, that can be discarded in a systematic renormalization 
of the mass of the boundary plate.
 (Formally, this quantity is the total energy of an isolated 
surface in otherwise empty space \cite{A}.)
 Again we consider for simplicity only the case where both 
horizontal boundaries are the same type, so that $(-1)^\eta$ is 
independent of~$k$.
 As a reminder that this assumption is in force, we shall write
 in resulting formulas
 \begin{equation}
 (-1)^\eta = \mp \equiv \cases{
 -1, &Dirichlet, \cr
 +1, &Neumann.  \cr}
 \label{DNsigns} \end{equation}
 The terms combine easily (telescope):
 \[ \sum_{k=-\infty}^\infty
 \int_0^a dx \int_0^b dy\, T_{\mathrm{V}0k}(t,\mathbf{r},\mathbf{r}) =
\mp \frac{at}{\pi} 
 \int_0^\infty
 \left(t^2  + 4y^2\right)^{-3/2} \,dy
 = \mp\, \frac a{2\pi t}\,. \]

  Obviously the formula for class HP is the same with $a$ 
replaced by $b$.            
 Therefore, the total contribution from VP and HP to the 
 trace of the cylinder kernel can be written as
 \begin{equation}
T_\bot(t)= \mp\, \frac P{4\pi t}\,,
 \label{perimcyl}\end{equation}
 where $P$ is the perimeter of the rectangle.
 It corresponds to a divergent surface energy
 \begin{equation}
 \langle E\rangle_{t\bot} = \mp\, \frac P{8\pi t^2}\,.
 \label{perimen}\end{equation}

For paths of class VD we obtain
 \[ \sum_{k=-\infty}^\infty 
 \int_0^a dx \int_0^b dy\, T_{\mathrm{V}jk}(t,\mathbf{r},\mathbf{r})
= \mp \frac{at}{2\pi} \,\frac1{t^2+(2ja)^2} 
 = \mp \frac{t}{8\pi aj^2} + O\left( \frac{t^3}{j^4} \right). \]
The sum over $j$ gives the well known $\zeta(2)$, so
 \[ \sum_{j\ne0} \sum_{k=-\infty}^\infty 
 \int_0^a dx \int_0^b dy\, T_{\mathrm{V}jk}(t,\mathbf{r},\mathbf{r}) =
\mp\, \frac{\pi t}{24a} + O\left( t^3\right). \]
 The corresponding contribution to the energy is
 $\pm \pi/48a\,$;
 it may be thought of as a Casimir correction to the surface energy 
of the sides at $y=0$ and $y=b$
  caused by the presence of the perpendicular 
sides with separation~$a$.
 Thus the total energy from VD and HD paths is (at $t=0$)
 \begin{equation}   \langle E\rangle_{t\mathrm{D}} =
    \pm\,\frac{\pi}{48}\left(\frac 1a + \frac 1b \right)  .
 \label{edgeenergy}\end{equation}
 It is comparable in magnitude to the  term 
 from periodic paths, (\ref{perenergy}).
In fact, for the square it is larger,  since $\pi/24 \approx 0.13$;
 that is why the ``renormalized'' (Lukosz) energy  
 of the Dirichlet square comes out positive.
 (The situation for the \emph{force} is different, however, as we 
shall see.)

 Finally, for a path of class C we have from (\ref{Cjk})
 \begin{equation}
 T_{\mathrm{C}jk}(t,\mathbf{r},\mathbf{r}) =
  \frac{(-1)^\eta}{2\pi} \,t
 \left[t^2 + (2ja-2x)^2 + (2kb-2y)^2\right]^{-3/2}.
 \label{Ccyl}\end{equation}
 Terms with $\{j,k\}\subset \{0,1\}$
  yield divergent integrals in the energy
(\negthinspace${}\sim \int r^{-3}r\,dr$) 
  if $t$ is set equal  to~$0$,
 but if one integrates with $t$ positive, the result is quite 
different.
  We assume that all sides are of the same type, so that
 $(-1)^\eta =+1$.
Then the contribution to the cylinder trace from the corner 
paths telescopes to
\begin{eqnarray}
\sum_{j=-\infty}^\infty \sum_{k=-\infty}^\infty 
 \int_0^a dx \int_0^b dy\, 
T_{\mathrm{C}jk}(t,\mathbf{r},\mathbf{r})
  &=&  \frac{2t}{\pi}  \int_0^\infty dx \int_0^\infty dy\, 
 (t^2+4x^2+4y^2)^{-3/2} \nonumber \\
 &=&\frac14\,.
\label{cornercyl} \end{eqnarray}                         
 Being independent of $t$, this term
  makes no contribution at all to the energy via~(\ref{00calc}).
 (In a related independent calculation by Zaheer et al.~\cite{Z}
the corner paths were not even considered, because the rectangle 
was obtained as a limiting case of a configuration where they did 
not exist.)
In the next subsection we shall review why this result is exactly 
what should have been expected.
  
\subsection{Relation to heat kernel asymptotics}
 \label{ssec:asymp}

 Let $K(t,\mathbf{r},\mathbf{r}')$ be the heat kernel corresponding 
to the system under study
 ($\langle \mathbf{r}|e^{-tH}|\mathbf{r}'\rangle$
  in quantum-mechanical notation, as contrasted with (\ref{cyl})).  
 Let $K(t)$ be its trace (cf.\ (\ref{ET})).
  It is well known \cite{Kac,Cl,Gil,Kir}
 that as $t\to0$ 
 \begin{equation}
 K(t) = \frac A{4\pi t} \mp \frac P{8\sqrt{\pi t}} +\frac14
 +O(t^\infty),
  \label{heatexp} \end{equation} 
 where $A=ab$ and $P = 2(a+b)$ are the area and perimeter of the 
rectangle, 
 and $\mp$ is as in (\ref{DNsigns}).
 (Here we state (\ref{heatexp}) only for the cases where 
 all four sides are of the same type. 
 The other cases
 --- in which, for instance, the second term is not 
proportional to  $P$, 
but the qualitative conclusions of this subsection remain true ---
  are discussed in~\cite{bookrev}.) 

 It follows \cite{systemat,BGH} that the trace of the 
\emph{cylinder} kernel must have the expansion 
 \begin{equation}
   T(t) = \frac A{2\pi t^2} \mp \frac P{4\pi t} + \frac14 + O(t),
 \label{cylexp}\end{equation}
 and hence by (\ref{ET}) the regularized Casimir energy is
  \begin{equation}
 E(t)\equiv -\,\frac12 \,\pd Tt 
 = \frac A{2\pi t^3} \mp \frac P{8\pi t^2} +
 \frac 0t + E_\mathrm{ren} +O(t),
 \label{enexp}\end{equation}      
 where $E_\mathrm{ren}$ is a constant traditionally identified as the 
 ``renormalized'' Casimir energy.  
 ($E_\mathrm{ren}$ is not determined by the 
 heat kernel expansion (\ref{heatexp});
 it is hidden in the $O(t^\infty)$ term there.)
 
 Our calculations above have confirmed (\ref{cylexp}) and 
(\ref{enexp})
 and determined   $E_\mathrm{ren}\,$.
 The terms in (\ref{cylexp}) are (\ref{Zterm}), (\ref{perimcyl}), 
 and (\ref{cornercyl}). 
  $E_\mathrm{ren}$ is the sum of (\ref{perenergy}) and (\ref{edgeenergy})
 (in agreement with  previous authors, including
  \cite{AW, Cav, Z}).

 With regard to the inevitability of the disappearance of the 
corner energy (without an explicit renormalization of any kind!),  
 we stress \cite{systemat,BGH} that the coefficient of $1/t$ in
 (\ref{enexp}) \emph{must} be $0$.
 (For dimensional reasons, that is where a corner term would need 
to appear, along with contributions linear in boundary curvature or 
in a Robin   constant.)
A $t^{-1}$ term in $E(t)$ would have to come from a $\,\ln t$ 
term in $T(t)$, which in turn would be associated with a 
logarithmic term in $K(t)$, and such terms do not exist.
 On the other hand, there is no general reason why $E(t)$ could not 
contain a $\,\ln t$ term (and a resulting scale ambiguity in the 
``renormalization''). 
 That would correspond to a $\,t\ln t$ in $T(t)$ 
and hence a $t^{1/2}$ in $K(t)$ ---  which can actually occur
 (for example in a disk), but 
does not in the model under study here.

 \subsection{Force and pressure calculations}
\label{ssec:force}

 We now investigate the force on the side at $x=a$ from the field 
vacuum inside the rectangle, in the case where all sides are 
Dirichlet.
 From the previous subsections, the naively renormalized 
 energy yields the force
 \begin{equation}
 -\, \pd{E_\mathrm{ren}}a =
 +\, \frac{\zeta(3)}{16\pi b^2}
 - \frac{\zeta(3)b}{8\pi a^3} 
 + \frac b{8\pi} \sum_{j,k=1}^\infty
      \frac{k^2b^2 - 2j^2a^2}{(j^2a^2 + k^2b^2)^{5/2} } 
 +\frac{\pi}{48a^2}
 +0,
  \label{renforce}\end{equation}
 where the terms are the contributions of path classes 
 PV, PH, PD, VD, and HD, respectively.
 
 It is important to remember that positive energy is not always the 
same thing as positive (repulsive) force, although that is true in
many of the classic Casimir-force calculations in which the 
absolute value  of the energy, being a negative power,
  decreases monotonically to $0$ as the relevant 
geometrical parameter increases.
 In (\ref{renforce}) the PV force is positive although the PV 
energy is negative; the individual terms in the PD force can have 
either sign, although their energies are all negative; the HD force 
is zero because the positive HD energy is independent of~$a$.
Piston analyses center on the cancellation of the positive PV
force by an external force, since the sum of the other three
terms can be shown to be negative (see Sec.~\ref{sec:piston}).

 As a first step toward less naive renormalization, 
  one can keep 
the cutoff $t$ finite and retain the cutoff-dependent terms in 
(\ref{enexp}).
 Then the VP term produces a force
 \begin{equation} F_{t\mathrm{VP}} =
 +\,\frac1{4\pi t^2}\,,
 \label{badforce}\end{equation}
 and the terms in (\ref{renforce}) are modified in ways that can 
become significant when $a$ or $b$ is not large compared to~$t$.

 Another way to calculate the force is to integrate 
  $\langle T_{11}(a,y)\rangle$  over the side of the  box.  
 We take a moment to verify that the methods are consistent, using 
the appropriate formulas from Sec.~\ref{ssec:Tjk}.

\emph{Periodic paths:} 
 Multiplying (\ref{T11Pjk}) by $b$ to perform the 
   trivial integration,
  we get (after $t\to0$)
\[ F_{\mathrm{P}jk} =\frac b{4\pi}\, 
 \frac{-2(2ja)^2 + (2kb)^2}{\left [(2ja)^2 
+(2kb)^2\right]^{5/2} }\,.  \]
Here $j$ and $k$ are not both $0\,$; the terms with one, the 
other, or neither $0$ add up to the first three terms in 
(\ref{renforce}), as expected.

\emph{Corner paths:}  From the energy calculation we know that 
these terms should be zero.
Also, from (\ref{T11Cjk}) we have $T_{11}=0$ unless 
 $\beta \equiv \xi-\frac14$  is nonzero.
For the $\beta$ term (\ref{dT11Cjk}), note that
 the integrand has the form of a total derivative, 
 \begin{equation}
 \frac {K -2(2bk-2y)^2 }{ [K +(2bk-2y)^2]^{5/2}} 
=  \od{}y\, \frac{y-bk}{ [K +(2bk-2y)^2]^{3/2}} \,.
 \label{Delcalc1}\end{equation}
Setting $x=a$ and integrating over $y$, one gets
 \begin{eqnarray}  F_{t\mathrm{C}jk} &=&
 -\,{\beta\over \pi}\, \left\{ \frac{(1-k)b}{
[t^2+4(j-1)^2 a^2 + 4b^2(k-1)^2 ]^{3/2}} \right. \nonumber \\
&&\qquad\left. {}
  -\frac{(-k)b}{[t^2+4(j-1)^2 a^2 + 4b^2k^2 ]^{3/2}} \right\}.
 \label{Delcalc2} \end{eqnarray}
 The sum of (\ref{Delcalc2}) over $k$ from $-\infty$ to $\infty$
telescopes to 0.

\begin{figure}
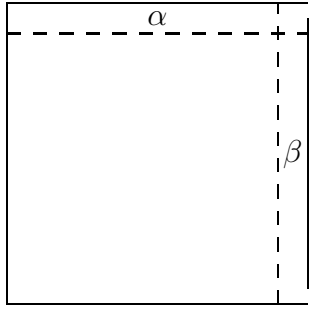

\centerline{\beginpicture
 \setcoordinatesystem units  <2truecm,2truecm>
 \putrule from 0 0 to 2 0
 \putrule from 0 0 to 0 2
 \putrule from 0 2 to 2 2
 \putrule from 2 0.1 to 2 1.9                             
 \setdashes
 \putrule from 0 1.8 to 2 1.8
 \putrule from 1.8 0 to 1.8 2
 \put{$\alpha$} at 1 1.9
 \put{$\beta$} at 1.9 1
 \endpicture}
 \caption  {Two regions where divergent surface energy appears.}
 \label{fig:alphabeta}\end{figure}

\emph{Side paths:}
The class of paths that bounce off the side in question (along 
with an even number of additional reflections) have $T_{11}$ 
identically zero
 (\ref{T12Hjk}).
   This is so even though the shortest such paths
 (those with $k=0$, $j=1$) give rise to a divergent energy in the 
region marked $\beta$ in Fig.~\ref{fig:alphabeta}.
    This matches the $0$ in~(\ref{renforce}).
  
 Much more interesting are the paths that bounce off the horizontal 
walls. 
 From (\ref{T11Vjk})--(\ref{dT11Vjk}), 
\begin{eqnarray*}
\langle T_{11}\rangle_{t\mathrm{V}jk}  &=& 
 -\,\frac1{4\pi} [t^2 +(2ja)^2 +(2kb-2y)^2]^{- 5/2} 
 [t^2 -2(2ja)^2 +(2kb-2y)^2] \\
 &&{}+\frac{\beta}{\pi} [t^2 +(2ja)^2 +(2kb-2y)^2]^{- 5/2}
  [t^2+(2ja)^2 -2(2kb-2y)^2] .
 \end{eqnarray*}
 For fixed $j,k$ the $\beta$ term is just like the corresponding 
corner term with $j-1$ replaced by $j$ and the sign changed.
 Therefore, these two classes of $\beta$ terms would cancel 
 when summed over~$j$, 
even if they did not vanish when summed over $k$ as we just saw.

 It remains to integrate the other 
 part of $\langle T_{11}\rangle_{t\mathrm{V}jk}$ over $y$ from 0 
to $b$.
 The terms with $j=0$ 
lead to a clone of the calculation following (\ref{Vcyl}).  
In particular, those terms for which also
$k=0$ or~$1$ are divergent when $t\to0$.
 This divergent pressure clearly corresponds,
in the case $k=1$, to the 
divergent energy in region $\alpha$  associated with  
 paths VP that bounce perpendicularly off the top boundary.
 (From $k=0$ comes a corresponding effect at the bottom boundary, 
not indicated in Fig.~\ref{fig:alphabeta}.)
That energy is  proportional to the 
length of the box and hence gives a force (\ref{badforce})
  upon differentiation.  

 Finally, one wants to integrate the terms with $j\ne0$ and see that 
they reproduce the remaining (VD) term in (\ref{renforce}).
    The integral of each term is, at $t=0$,
 \[    -\,\frac1{4\pi}\,
    \frac{\frac{b[32a^4j^4 +16a^2b^2j^2(k-1)^2](k-1)}
    {[4a^2j^2+4b^2(k-1)^2]^{3/2}}
    -\frac{bk[32a^4j^4 +16a^2b^2j^2k^2]}
    {[4a^2j^2+4b^2k^2]^{3/2}}
    }{ 16a^4 j^4 }\,. \]
At first glance it may seem that this expression sums over $k$ 
 to zero, by the same telescoping argument used elsewhere.
However, unlike those previous sums,
in this case the individual terms do not approach 0 as 
$|k|\to\infty$; rather, they go to $1/(32\pi a^2j^2)$.
 Taking account of both signs of $k$ and~$j$,
 one gets    the force to be
 \[F_\mathrm{VD}=
 4 \sum_{j=1}^\infty\frac1{32\pi a^2j^2} 
 = \frac{\zeta(2)}{8\pi a^2}
    = \frac{\pi}{48a^2} \,,\]
    as needed.

Although this exercise may appear redundant, it has underscored two 
important points.
 First, doing the calculation in terms of pressure instead of 
energy by no means eliminates the problem of divergences.
 Second, the divergent pressure on a given wall is not associated 
with the divergent energy adjacent to the wall 
 (in region $\beta$ in Fig.~\ref{fig:alphabeta}).
 Rather, it goes with the divergent energy adjacent to the 
intersecting perpendicular walls 
 (such as in region $\alpha$).

  \section{The Casimir piston} \label{sec:piston}

 The physical significance of the forces calculated in \cite{Lu}, in 
our Sec.~\ref{ssec:force}, and in much intervening literature has 
been called into question.
 For one thing, unlike the celebrated sphere calculations of Boyer 
\cite{Boy} and others, these calculations are unable to take into 
account any forces coming from the region outside the box.
 In addition, 
within the framework of ultraviolet-cutoff 
regularization the 
uncompensated divergent energy proportional to the surface area 
 cannot be easily dismissed in deducing the force conjugate 
to a dimension whose variation changes the surface area.
 In our case the offending energy is that localized in the region 
$\alpha$ in Fig.~\ref{fig:alphabeta},
  which is proportional to the length of the box,
  and the corresponding pressure was also observed in 
Sec.~\ref{ssec:force}  in the direct calculation of 
  $\langle T_{11}\rangle$ on the movable side of the box.

\begin{figure}
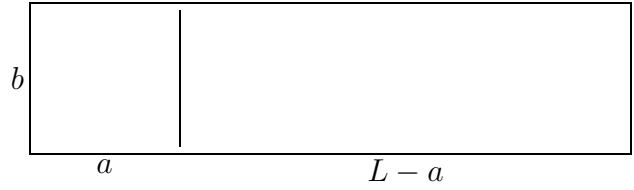

\centerline{\beginpicture
 \setcoordinatesystem units <2truecm,2truecm>
 \putrule from 0 0 to 4 0 
 \putrule from 0 1 to 4 1
 \putrule from 0 0 to 0 1
 \putrule from 1 0.05 to 1 0.95
 \putrule from 4 0 to 4 1
 \put{$b$} [r] <-2pt,0pt> at 0 0.5
  \put{$a$} [t] <0pt,-2pt> at 0.5 0
   \put{$L-a$} [t] <0pt,-2pt> at 2.5 0
 \endpicture}
 \caption{A rectangular piston in 
dimension~2.  Its ``shaft'' has length $L-a$, effectively infinite.
 The word ``piston'' refers both to the movable plate at $x=a$
  and to the model as a whole.} 
\label{fig:piston} \end{figure}

 Cavalcanti \cite{Cav} proposed to avoid both problems by 
considering a different situation, the piston 
(Fig.~\ref{fig:piston}).
The interior partition is free to move horizontally, and one is to 
calculate the force upon it.
 $L$ is to be taken very large compared to $a$ and~$b$.
The argument now is that the exterior of the apparatus is 
unchanging and hence irrelevant to the force, whereas both interior 
chambers can be treated exactly.
 Furthermore, the total of the interior side lengths is independent 
of the piston position, $a$, so that the surface divergences cancel 
in the calculation of the force.

 Generalizations and variations of this model have been extensively 
studied 
 \cite{Sv,HJ1,HJ2,Bar-pist,Mar,Ed-pist,ZL,SchM,Mar-pist,
EdMd,EdM,Cheng,Schaflask}.

 The piston model is not without its own physical problems, because 
interactions between the piston plate and the horizontal sides have 
been ignored.
  In a realistic experiment, because of ordinary Casimir 
attraction the plate 
 would be unstable to striking 
the tube wall on edge, after which
it would collapse against one 
of the walls of the tube.
   It may be argued that this
objection is irrelevant to the question of principle that the 
piston model is designed to address;
    the only degree of freedom one is varying is~$a$,
    so it is legitimate to imagine that the plate is constrained 
    from moving 
in any other degree of freedom.
    There still exists a Casimir force between the plate and the 
nearest wall, though it is somehow prevented from causing motion.
    However, one can argue by symmetry that
    this force has no significant horizontal component, so 
that the piston theorists are  justified in ignoring it.
   Nevertheless, in a real apparatus there would surely be some 
   friction with the walls, so the feasibility of an experiment to 
   verify the piston analysis is questionable.

 Putting these doubts aside, we summarize and recast the Cavalcanti 
analysis in our framework of closed paths.
 The finite part of the force on the piston from the chamber on the 
left has been calculated in~(\ref{renforce}).
 The force coming from the shaft on the right can be found from the 
same formula, with the sign reversed, $a$ replaced by $L-a$, and 
$L$ taken to infinity; the only term that survives is the PV term,
 \begin{equation}
  F_L = -\,{\zeta(3)\over 16\pi b^2}\,.
 \label{shaftforce}\end{equation} 
It exactly cancels the corresponding term in~(\ref{renforce}),
 leaving PH, PD, and VD terms:
 \begin{equation}
 F_\mathrm{pist} = 
 -\, \frac{\zeta(3)b}{8\pi a^3} 
 + \frac b{8\pi} \sum_{j,k=1}^\infty
      \frac{k^2b^2 - 2j^2a^2}{(j^2a^2 + k^2b^2)^{5/2} } 
 +\frac{\pi}{48a^2}\,.
 \label{pistforce}\end{equation}
 Here there is no ``naive renormalization'' 
 as in (\ref{renforce}),
  since the divergences (in particular, the VP terms) would 
  explicitly cancel if the calculation were done for the complete 
  piston before removing the cutoff.

 Cavalcanti \cite{Cav} rendered (\ref{pistforce})
more illuminating by
 subjecting it to  further analysis.
 If one refrains from the $\zeta(3)$ simplification, the 
 complete sum over periodic paths in the ($t=0$) energy, 
 (\ref{perenergy}),  is
 \begin{equation}
 \langle E\rangle_{t\mathrm{P}\setminus\mathrm{Z}} =
   -\frac{ab}{32\pi} \sum_{j,k=-\infty \atop (j,k)\ne(0,0)}^\infty
   (j^2a^2 + k^2b^2)^{-3/2}.
 \label{penergy}\end{equation}
 From this one can derive two complementary formulas, useful in the 
respective regimes $a\gg b$ and $a\ll b$.
 (Unfortunately, none of the three formulas for $F_\mathrm{pist}$ 
 is  completely transparent for $a\approx b$.)

In the first case, 
  for $j=0$ one  evaluates the $k$ sum to the term~PV
 (the first term in~(\ref{perenergy})), as 
before, but
 for  fixed $j\ne0$, one applies a known relation between the $k$~sum 
 (which is an Epstein zeta function)  and a series of modified 
Bessel functions.
(This theorem traces back ultimately to the Poisson summation 
formula;
see the appendices of \cite{AW} and \cite{Kir}.) 
  Thus the PH and PD terms together are replaced by the energy 
  terms
 $$-\,\frac{\pi}{48a} - \frac1{2b} \sum_{j,k=1}^\infty \frac 
kj\,K_1\left(2\pi jk\,\frac ab \right).$$
(The individual terms in the sum cannot be associated with 
individual periodic orbits, nor with individual eigenvalues.)
 Remarkably, the first term of this expression precisely cancels 
the VD term, so that PH, PD, and VD all together reduce to the 
energy
 $$- \,\frac1{2b} \sum_{j,k=1}^\infty \frac kj\,
 K_1\left(2\pi jk\,\frac ab \right).$$
 Since HD does not contribute to the force and the PV force 
 is still 
cancelled by the force from the shaft, the force on the piston is 
\cite[(11)]{Cav}
 \begin{equation}
 F_\mathrm{pist} =\frac{\pi}{b^2} \sum_{j,k=1}^\infty k^2
  K_1'\left(2\pi jk\,\frac ab \right).
 \label{fbiga}\end{equation}
It follows that the piston force (a) is always negative,
 (b) vanishes exponentially fast for $a\gg b$, in contrast to the 
usual power-law decay of the Casimir force.

 Alternatively, one can apply the Epstein-to-Bessel
 transformation to the $j$ sum for fixed~$k$.
 That is, PV and PD get replaced by
 $$-\,\frac{\pi}{48b} - \frac1{2a} \sum_{j,k=1}^\infty \frac 
kj\,K_1\left(2\pi jk\,\frac ba \right).$$
The first term cancels HD (which doesn't contribute to the force 
anyway);
 the VD term remains (as does PH); and the PV term has been 
absorbed, so that 
the force from outside the piston is now uncompensated.
 Thus the total force on the piston is \cite[(14)]{Cav}
 \begin{equation}
 F_\mathrm{pist}=
 -\, {\zeta(3)b\over 8\pi a^3} + \frac{\pi}{48a^2} 
 -\frac{\zeta(3)}{16\pi b^2} +\frac{\pi b}{a^3} 
 \sum_{j,k=1}^\infty
 k^2 K_0\left( 2\pi jk\, \frac ba\right).
 \label{fsmalla}\end{equation}
 As Cavalcanti explains, this form is nicely adapted to 
understanding the regime $a\ll b$, where the standard Casimir 
result is, of course, recovered in the limit.

Energy is fungible, so one must beware of attributing too 
fundamental a connection between particular classes of paths and 
the observable net forces.
 The striking thing is that the calculations reveal several exact 
cancellations, not all of which can be implemented at  the same 
time.

\section{The Casimir pistol}
\label{sec:pistol-d}

\begin{figure}
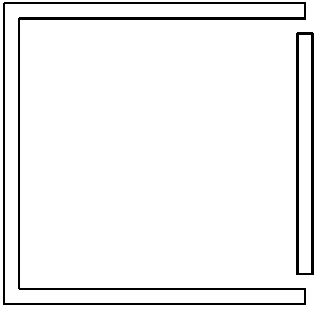

 \centerline{\beginpicture
 \setcoordinatesystem units  <2truecm,2truecm>
 \putrule from 0 0 to 2 0
 \putrule from 0 0 to 0 2
 \putrule from 0 2 to 2 2
 \putrule from 1.95 0.2  to 1.95 1.8                             
 \putrule from 0.1 0.1 to 2 0.1
 \putrule from 0.1 0.1 to 0.1 1.9
 \putrule from 0.1 1.9 to 2 1.9
 \putrule from 2.05 0.2 to 2.05 1.8
 \putrule from 1.95 0.2 to 2.05 0.2
 \putrule from 1.95 1.8 to 2.05 1.8                              
 \putrule from 2 0 to 2 0.1 
 \putrule from 2 2 to 2 1.9 
 \endpicture}
 \caption{A rectangular box with one side (``lid'') free to move.
 The box has walls of finite width 
and a finite gap between the lid and the box sides.}
 \label{fig:looselid}\end{figure}

The Casimir piston has proved to be a highly illuminating model, 
but it does not settle the issue of the true physical significance 
of the purely internal vacuum pressure on the side of a rectangular 
cavity.
 In the piston both the divergent (VP) 
 and the positive finite (PV)
 internal  pressure are exactly 
balanced by the precisely analogous pressures in the long shaft on 
the other side of the movable plate.  This observation does not 
tell us what would happen if the external shaft were not there.
The problem of real interest is  
  a rectangular box with one side free to move, as indicated 
  schematically in  Fig.~\ref{fig:alphabeta} and more realistically 
  in Fig.~\ref{fig:looselid}.
     The main question is whether the force on the 
movable side is attractive or repulsive.  This is a question 
about disjoint rigid bodies, so it is a 
meaningful physical problem, just like the piston.
    (The   3D electromagnetic case 
should be qualitatively similar to the 2D scalar problem.)     
 Another urgent question is what happens to the VP divergence now 
that there are no VP paths in the shaft to compensate it;
the previous paradox (Sec.~\ref{sec:global})
of an apparent infinite pressure has reappeared.

    The problem is difficult because there is no reliable 
analytical calculation of
 the forces acting from outside the box and inside the tiny gaps at 
the ends of the lid. 
If we momentarily ignore the gaps,   it seems unlikely 
that the external forces would be very large
 (although we find unconvincing 
 Lukosz's attempt \cite{Lu} to prove this fact by 
appealing to Weyl's theorem).
 If one thinks in terms of closed paths, paths striking the walls 
perpendicularly will yield only the usual surface divergence, so 
the only possible source  of nontrivial external forces
is the diffractive paths striking the corners.
 If this diffractive effect is small,
therefore,  one might expect the force to be repulsive when the 
plate is exactly at the mouth of the box.  However, 
if the plate is located significantly inside or outside the box, 
 intuition says the opposite: ``inside'' we are getting into the 
 piston regime, 
    whereas  ``outside'' the Casimir attraction between the 
nearest neighboring regions of the two bodies should be dominant.

 A convincing resolution of this apparent paradox presumably 
requires a serious study of the gap region in a less idealized 
geometry, as in Fig.~\ref{fig:looselid}.
It is 
clear that what happens around the gap is very complicated, 
especially when the plate is part-in and part-out as in that 
figure.
 One should  note  that the symmetry argument used in 
Sec.~\ref{sec:piston} to dismiss the forces in the gap is
   no longer applicable.

\begin{figure}
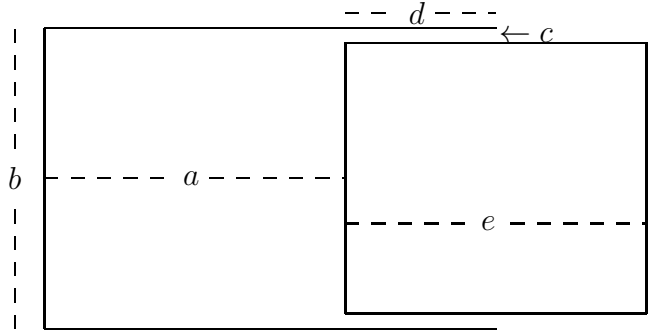

 \centerline{\beginpicture
 \setcoordinatesystem units  <2truecm,2truecm>
 \putrule from -1 0 to 2 0
 \putrule from -1 0 to -1 2
 \putrule from -1 2 to 2 2
 \putrule from 1 1.9 to 3 1.9                             
 \putrule from 1 0.1 to 3 0.1                             
 \putrule from 1 0.1 to 1 1.9
 \putrule from 3 0.1 to 3 1.9
 \setdashes
 \putrule from -1 1 to -0.15 1                     
 \putrule from 0.1 1 to 1 1
 \put{$a$} at -0.025 1 
  \putrule from -1.2 0 to -1.2 0.85
 \putrule from -1.2 1.2 to -1.2 2              
 \put{$b$} at -1.2 1
 \put{$\leftarrow c$} at 2.2 1.95
 \putrule from 1 2.1 to 1.35 2.1
 \putrule from 1.6 2.1 to 2 2.1
 \put{$d$} at 1.475 2.1
 \putrule from 1 0.7 to 1.85 0.7
 \putrule from 2.1 0.7 to 3 0.7
 \put{$e$} at 1.95 0.7
 \endpicture}
 \caption{The Casimir pistol, consisting of two disjoint, perfectly 
conducting bodies, the barrel and the bullet.
 It is shown schematically, as in Fig.~\ref{fig:alphabeta},
but the barrel can be thought of as having finite thickness, as in
 Fig.~\ref{fig:looselid}.}
 \label{fig:pistol}\end{figure}

The uncertainty about the external and gap forces is somewhat 
alleviated 
    if we replace the thin lid 
by a large rectangular object (Fig.~\ref{fig:pistol}).
   The piston plate has now become more 
like a bullet or artillery shell.  The 
question now is the sign of the force for various values of the 
five dimensions indicated in Fig.~\ref{fig:pistol}: 
    Does a Casimir pistol exist?
The advantage of this new problem is that the corners of the 
two bodies are not near each other, so there are no short 
classical paths outside the apparatus
 (as long as neither $d$ nor $e-d$ is small), even if diffractive paths
are admitted as classical.
(One could eliminate diffractive paths
  (in the sense we are using the term) by replacing the barrel and 
bullet by similarly shaped objects with smooth boundaries.)
  Like all piston authors, we continue to consider only 
  horizontal motion (variation of~$a$) and therefore ignore the 
  vertical force between the bullet and the shaft of the barrel.

    We now consider the implications of taking the small gap of 
width $c$ seriously. 
    (We speak only of the gap at the top, but obviously the same 
remarks apply to the one at the bottom.) 
    The first (and motivating) observation is that the total side 
length of the system is now fixed, and hence so is the transverse 
extent of the infinite (or cutoff-dependent) surface energy.
 In particular, the energy associated with what we call VP paths 
 (including those striking the exterior of the apparatus) is 
independent of~$a$.
    The associated paradox  is thereby  removed!

 Our joy in this victory should be short-lived,
    however.
  If we take VP paths inside the gap 
seriously, then for consistency we must also take PV
 paths across the gap seriously, and we shall see that they 
present a serious problem.

    Nonperpendicular paths inside the gap eventually escape from 
it, so they are not ``short'' and probably can be neglected.
 (In the model of \cite{Z}, all such paths escape to infinity and 
hence can never be closed.  In our case it is possible, but rare,
  for such a 
path to bounce off the left side of the rectangle and return to its 
starting point.) 
 From the point of view of a point $\mathbf{r}$ inside
     the rectangular region of area $ab$,
 the box now has 
 small ``leaks'' of  width~$c$,
  but one would not expect that to affect its 
internal Casimir energy significantly.
 This observation could be made quantitative by imitating a 
calculation in~\cite{Z}, but we shall not do so here, because we 
are interested only in the limit of very small~$c$.

    On the other hand, because the gap $c$ is much
smaller than the box dimensions, $a$ and $b$, the Casimir energy 
associated with the rectangle of area $cd$ is much 
greater than that of the box.
The principal force associated with this gap rectangle is the 
vertical Casimir attraction between the bullet and the barrel, 
    but we have 
agreed to impose a constraint that makes it irrelevant.
    However, it is the very essence of the piston argument, 
especially as developed by Hertzberg et al.~\cite{HJ1,HJ2},
    that the proportionality of the Casimir energy to~$d$ 
produces a horizontal force, independent of~$d$ but proportional 
to $1/c^2$.
        (This energy is precisely the contribution of the 
vertical periodic (PV) paths.)
    In the present scenario this force has sign opposite to the 
Lukosz force in the box, because $d$ increases when $a$ 
decreases, and a larger magnitude than the analogous force in the 
piston scenario, because $c<b$. 
Therefore, if we accept all the approximations 
involved in this argument, we are forced to the conclusion that 
the bullet is sucked into the barrel, not expelled from it.

Let us list those assumptions. 
    \begin{enumerate}
    \item The gap between the bullet and the barrel does not 
significantly affect the Lukosz force from the empty part of the 
pistol chamber.
    \smallskip
    \item There are no significant forces from outside the pistol.
    \smallskip
    \item The effect of the gap can be estimated by ignoring 
nonperpendicular paths and treating the perpendicular paths as 
usual, as if we had simply a pair of parallel plates there.
 \end{enumerate}

Obviously, a trustworthy treatment of this system 
 requires either a numerical analysis 
(for example, by the method of Gies et al.~\cite{GK}) 
 or, better still, a 
 breakthough in the analytical treatment of convex corners.
As a positive result in this direction, we report that
 it is possible to compute exact forces and 
torques between bodies of arbitrary shape in weak coupling 
(for example,   
materials with dielectric constant nearly unity).  For example, 
two thin parallel plates of finite length experience an 
attractive lateral force that tends to cause the plates to move 
to a configuration where they are centered on each other.  This 
is the attractive force that tends to increase the length of the 
gap in the pistol.
(A recent independent investigation \cite{RJJ} likewise shows a
system maximizing the length of a small gap between flat
surfaces.)
 Moreover, in 
addition to the attractive force between the plates, there is a 
torque exerted on one thin plate above a larger plate which tends 
to cause a rotation of the smaller plate about its center of mass 
so as to favor perpendicular orientation.  Details 
have been reported in~\cite{MPW,MPWconfs}.  Since these 
qualitative conclusions are essentially geometrical, they should 
also hold for strong coupling (Dirichlet boundary conditions).

 Finally, let us try to confirm the foregoing conclusions   
  by    looking at pressure integrals.
 In principle, one can find   
  the total force on the bullet by integrating 
the appropriate components of $\langle T_{\mu\nu}\rangle$
  over the surface of the bullet
 (or even some larger surrounding surface \cite{R1,R2}).
     The integral over the back 
side of the bullet is essentially the same as in Sec.~\ref{sec:global}
  (apart from the ``infinite'' term).
   On the top and bottom sides, the 
relevant component is $T_{12}\,$, and a check of the 
formulas (\ref{T12Pjk}), (\ref{T11Cjk}), (\ref{T22Vjk}), (\ref{T12Hjk})
 shows that the contributions all vanish.
 So, one would conclude that the pistol fires after all!
    We believe that the resolution of this new paradox is that the 
crude approximations listed above, although they 
\emph{may} be permissible for the energy calculation, are simply 
wrong for the pressure calculation.  In particular, if the 
Casimir (or the Lukosz) formulas were accurate over the entire 
gap rectangle, there would be  finite jumps at the end surfaces
of the gap in $\langle T_{11}\rangle$ 
 (which is constant and large in the gap, 
constant and smaller in the chamber, and zero in the exterior, in 
our approximations).  By the conservation law, there is then a 
delta function (of $x- \mathrm{(endpoint)}$) in
     $\langle \partial T_{12} /\partial y\rangle$.
    A more realistic calculation would smear out this 
singularity, probably creating a lump of $\langle T_{12}\rangle$ 
 that decreases 
more or less linearly in $y$ away from the  back corner of the 
bullet and also downward away from the front edge of the barrel.
 These stress terms would create horizontal forces.
  They are very much like the stresses found in 
 \cite[Figures 4(d,e,f)]{R1}.

 \section{The Casimir pistol with cutoff} \label{sec:pistol-c}

 \subsection{Parallel plates revisited}

Our discussion so far has concerned the 2D scalar analogue of the 
idealized perfect-conductor model of the interaction of the 
electromagnetic field with metal bodies.
 It is generally agreed that the divergences (except for the 
universal volume divergence) encountered in such calculations are 
the fault of the physical failure of that model at high 
frequencies
 --- equivalently, at length scales so small that the material 
cannot be modeled as a continuum.
 It is also now agreed that 
 the energy divergences, or  the corresponding cutoff-dependent 
terms in a calculation with a cutoff, being independent of the 
bodies' positions, do not appear in the forces between
 rigid conducting bodies.
 It is sometimes forgotten that the idealized Casimir theory runs 
into physical trouble already for rigid bodies, even the canonical 
scenario of parallel flat plates, when the distances become too 
small.  
 It predicts an energy per unit cross section, $\mathcal{E}$,
  proportional to 
 $-a^{-d}$ for plates with separation $a$ in $d$-dimensional space.
 If taken literally, this says, implausibly,
  that $\mathcal{E}$ becomes (negatively)
  infinite when $a$ goes to zero.
 One would expect instead that in that limit $\mathcal{E}$ 
 approaches a constant,
 since then the space between the plates has disappeared and space 
is filled by the perfectly conducting material.
(In fact, the constant should make the total energy turn out to
be $0$ when suitably defined surface energies are also taken into
account.)

  Barton \cite{Bar} has done extensive calculations for 
  dielectric bodies with a polarizability small enough to be 
  treated perturbatively
 (the opposite regime from perfect conductivity). 
He showed (see also  \cite{maraball}) that a 
  spatial cutoff at atomic distances serves to cure the 
divergences (which otherwise remain even in
  the usual model of 
  quadratic falloff of dielectric constant with frequency
--- e.g., \cite{BE,MN}).         
 Roughly speaking, the mathematical effect of such a 
  cutoff is similar to that of a very rapid, such as exponential,
 cutoff at high   frequency.
 In Barton's theory the total energy per unit cross section does 
approach $0$ as $a\to0$ when the surface energy is included.
    In this model, as $a\to0$ there is a constant attractive 
force proportional to the energy density of the uniform medium, 
no matter how the latter is regulated \cite{milnotes}.

    More recently, Barton \cite{Bar-sphere,Bar-sheet} has 
developed a plasma model that is more pertinent to the limit of 
perfect conductivity.  It also involves an atomic-scale cutoff, 
but one affecting only the wavelengths parallel to the 
boundaries.

 Our aim here is to stay in the highly conducting regime and 
 to see whether keeping the exponential cutoff parameter~$t^{-1}$ 
 finite, at some value typical of atomic separations, yields a 
physically plausible (and divergence-free) model of Casimir 
phenomena. 
 Although ultimately no substitute for serious microscopic modeling 
of conductive materials (an unavoidably nonlinear problem), 
 this approach offers hope of rescuing the huge investment that has 
been made into treating vacuum problems (relatively easily) by 
spectral analysis of linear partial differential operators.
 It also provides a route to understanding the gravitational 
significance of ``divergent'' local energies and stresses
 \cite{leipzig}. 
    This cutoff should be regarded as analogous to the ad hoc 
repulsive core in the Lennard--Jones potential in atomic physics.  
A more accurate potential should be based on the electronic 
structure of the atoms; but one would not then apply such a 
potential to, say, nucleon-nucleon scattering.  Similarly, a 
detailed theory of real metals is not relevant to hadron bags, 
cosmological branes, thermal fluctuations in soft-matter physics, 
and other systems where Casimir-like effects have been studied.  
Within our two-dimensional scalar model (which is pertinent to 
all these contexts, if to any) the simple exponential cutoff has 
the advantage of being universal, but we and readers must remain 
conscious that its relevance at small distances to any particular 
real physical system is qualitative at best.  We stress again that 
this atomic-scale cutoff must not be confused with the well known 
decrease of dielectric constant with frequency above the plasma 
frequency; we make no attempt to model the latter, which is 
specific to the electromagnetic scenario.

In the context of the Casimir pistol, the idea is that the small 
gap $c$ surrounding the bullet must be in the sub-Casimir regime if 
the other dimensions ($a$, $b$, $d$) are in the regime where 
Casimir effects are significant, and,
  therefore, the deduction in Sec.~\ref{sec:pistol-d}
 of a dominant attractive force originating in the gap goes outside 
the regime of validity of the theory.
 Although the cutoff theory has no fundamental physical 
justification, it is probably a bit closer to the truth.
 (Unfortunately, we shall see that no robust conclusion is 
attainable by this route.)

 Maclay and Villarreal \cite{MV} proposed this same kind of 
     cutoff and 
hence obtained formulas and graphs rather similar to ours in this 
section.  However, they identified $t$ with the reciprocal of the 
plasma frequency rather than, as we do, the interatomic spacing, 
which is typically 100 times smaller (again cf.~\cite{Bar}).
    Other authors \cite{MPS,perivo} considered (for refutation) 
an even bigger exponential cutoff length, adequate to make the 
unrenormalized vacuum energy of empty space consistent with 
cosmological observations, and showed that such theories predict 
Casimir repulsion at distances large enough to be refuted by 
the existing laboratory experiments.

 Before studying the pistol, let us look at the 
attraction between parallel plates with the cutoff retained.
 This could be done easily in any dimension, 
but for coherence  in this paper we retain dimension~$2$.  
 If the separation between plates is $a$, we take the $b \gg a$
limit of (\ref{regperenergy}), 
in which only the perpendicular paths ($k = 0$)
contribute, and divide by $b$ to get energy per unit length:
 \begin{equation}
 \mathcal{E} =                 \frac{a}{\pi} \sum_{j=1}^\infty
\frac{t^2 - 2j^2a^2}{(t^2+ 4j^2a^2)^{5/2}}\, .
 \label{plateenergy} \end{equation}
This function  behaves
 in keeping with the idea that the energy or the force should
be damped when $a$ is comparable to the nanoscale (interatomic 
spacing) represented by~$t$.
 Of course, when $a\gg t$, the effect of $t$ is negligible and 
(\ref{plateenergy}) gives the standard result.
  It is convenient to measure $a$ in units of~$t$. If
 $ a = st$, then $\pi t^2\mathcal{E} = F(s)$, where
 \begin{equation}
  F(r) \equiv    r \sum_{j=1}^\infty
 \frac{1  - 2j^2r^2}{(1+ 4j^2r^2)^{5/2}} 
 \label{platedimenless}\end{equation}
 \begin{figure}
\centering \includegraphics{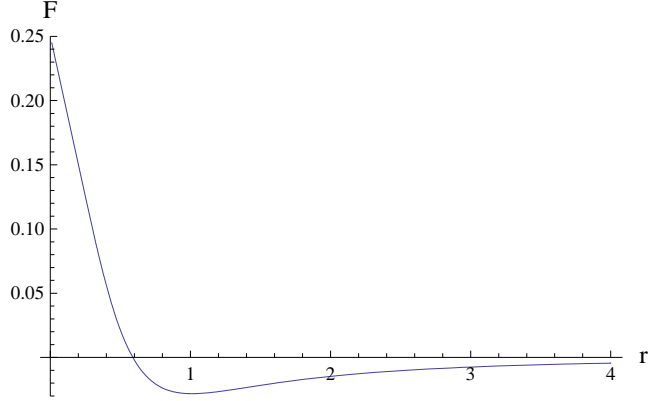}
 \caption{Graph of $F(r)$ as a function of $r$.}
 \label{fig:plates2d}
 \end{figure}
(see Fig.~\ref{fig:plates2d}).
 $F(r)$ has
  a zero at $r_0 \approx 0.5888$. It has a minimum
 (a zero of the force) at $r_1 \approx 1.0105$, 
 with $F(r_1)\approx -0.02821$.
 At large $r$, $F(r)\sim -\zeta(3)/16r^2$ as in the theory without 
cutoff.
 For small $r$ the Euler--Maclaurin formula \cite[(23.1.30)]{AS}
 shows that
 $F(r) \sim \frac14 - \frac r2 +O(r^N)$
 for arbitrarily large~$N$.                           
Thus $F(0)$ precisely cancels the surface energy from (\ref{perimen})
 (where $P=2$ because we are looking at unit cross section on two 
plates), so that the total energy at $s=0$ is indeed~$0$.
 (But this result may be an accident.  It does not happen for the 
Neumann boundary condition.  Also, as we shall now 
observe, $s=0$ seems to represent material under compression, not a 
solid block of ordinary material.)

At $r<r_1$ this model predicts a repulsion.  Therefore, it must 
violate the hypotheses of the theorems stating that vacuum forces 
between  symmetrical bodies separated by a plane are always attractive
 \cite{KK,Bac}.
 The argument of Kenneth and Klich \cite{KK} refers to the standard 
dielectric model of the media, 
or a scalar analog thereof,
into which our cutoff does not fit.
The mathematical reason why the theorem of Bachas \cite{Bac} 
doesn't apply is less clear, but the key physical point is clear 
from that author's remarks (p.~9094) 
 that a ``quantized particle does not, 
strictly speaking, live in one side of the reflecting plane,''
and that the theorem would apply at the quantum level
only if (in the terms of our scenario) one of the slabs were made
of antimatter.
 The repulsion occurs only at separations of the order of the 
interatomic spacing.  Thus the model mocks up a more realistic 
theory in which the two  slabs are not cleanly separated, in 
accordance with Barton's remark \cite[p.~4088]{Bar},
 ``[A sharp short-distance cutoff,] though a fiction, is a 
convenient shortcut to somewhere near the truth.
 At small separations, overlap between the electron clouds makes 
the interatomic potential highly repulsive....''
 We recall also that Ford and Svaiter \cite{FS} 
found that a similar effect was induced by a 
 stochastic uncertainty in the position of the conducting 
boundaries, which must in general lead to some probability  of 
 interpenetration.
 In short, nobody should be surprised to encounter a repulsion when 
pushing two slabs of material together.
    A normal, stable material must resist compression.  Of 
course, such repulsion is not a ``Casimir effect'';
a quantitative
study would require detailed modeling of the material, and it is 
a surprise and probably an accident that our crude
field-theoretic model gives such plausible results in this regime.  
In particular, the fact that our potential minimum occurs at 
a small positive separation, rather than zero or negative, is not 
to be taken too seriously.

 \subsection{Energy in the pistol}

 Now we do the energy accounting for the pistol, under the three 
assumptions listed in Sec.~\ref{sec:pistol-d}. 
 (The notation is a slight simplification of that
  in Sec.~\ref{sec:global}.)

 \emph{Energy in the chamber:} According to assumption~(1),
the contribution $E_\mathrm{P}$ of periodic paths is still given by 
 (\ref{regperenergy}) with $\eta=0$.
 Corner paths can be ignored because they make no contribution to 
the total energy. 
HP and HD paths 
 can be ignored here because they make no contribution to the 
relevant force (their energies being independent of~$a$).
The contribution of VP paths is the $a$-dependent term of 
 (\ref{perimen}):
 \begin{equation}
 E_\mathrm{VP}  =-\, \frac a{4\pi t^2}\,.
 \label{Ev}\end{equation} 
The contribution of VD paths is given by the generalization of
(\ref{edgeenergy}) to finite cutoff,
\begin{equation}
 E_\mathrm{VD} = \frac a{2\pi}   \sum_{j=1}^\infty 
{-t^2+4j^2a^2 \over (t^2+4j^2a^2)^2 }\, .
 \label{Ed}\end{equation}

 \emph{Energy in the barrel:}
 In the formulas above,  $(a,b)$  must be replaced  by $(d,c)$,
 and we must multiply by $2$ to count both top and bottom gaps.
 In accordance with assumption (3), only vertical paths ($j=0$) will be 
considered.
The PV paths give
 \begin{equation}
  E_{\mathrm{P}'} = \frac{2cd} {\pi} \sum_{k=1}^\infty
 {t^2 - 2k^2c^2  \over (t^2 + 4k^2c^2)^{5/2} }\,. 
 \label{Epprime}\end{equation}
The VP paths give
\begin{equation}
  E_{\mathrm{VP}'} = -\, \frac d{2\pi t^2}\,,
 \label{Evprime}\end{equation}
 of which half belongs to the barrel and half to the bullet.
 As expected, the barrel part combines with (\ref{Ev}),
 \[E_\mathrm{VP}+\frac12 E_{\mathrm{VP}'} = -\,\frac{a+d}{4\pi t^2}\,,\]
 to yield something independent of $a$, because $a+d$ is constant.
 Similarly, the bullet part of (\ref{Evprime}) combines with the 
surface energy of the part of the bullet outside the barrel.

 \emph{Summary of pistol energy:}
 The only energy terms that contribute to the force (under our 
assumptions) are
 \begin{equation}
  E= E_\mathrm{P} + E_\mathrm{VD} + E_{\mathrm{P}'}
 \label{Etot}\end{equation}
 as listed above.
 We could differentiate with respect to $-a$ 
 (using $\pd {}d = - \pd{}a$) to get the force.
  All the sums 
encountered can be expressed in terms of  inhomogeneous Epstein 
zeta functions \cite{El,Kir}.
However, for our purposes
 it is  better to analyze the various terms qualitatively. 
 (Quantitatively, we claim nothing for the model at short distances 
 anyway.)

  \subsection{Asymptotics and numerics for the pistol}

\begin{figure}
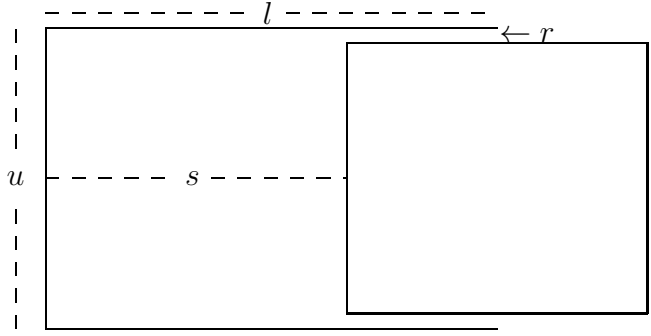

 \centerline{\beginpicture
 \setcoordinatesystem units  <2truecm,2truecm>
 \putrule from -1 0 to 2 0
 \putrule from -1 0 to -1 2
 \putrule from -1 2 to 2 2
 \putrule from 1 1.9 to 3 1.9                             
 \putrule from 1 0.1 to 3 0.1                             
 \putrule from 1 0.1 to 1 1.9
 \putrule from 3 0.1 to 3 1.9
 \setdashes
 \putrule from -1 1 to -0.15 1                     
 \putrule from 0.1 1 to 1 1
 \put{$s$} at -.025 1 
  \putrule from -1.2 0 to -1.2 0.85
 \putrule from -1.2 1.2 to -1.2 2              
 \put{$u$} at -1.2 1
 \put{$\leftarrow r$} at 2.2 1.95
 \putrule from -1 2.1 to 0.4 2.1
 \putrule from 0.6 2.1 to 2 2.1
 \put{$l$} at 0.475 2.1
 \endpicture}
 \caption{Pistol dimensions in units of $t$.}
 \label{fig:pistoldimenless}\end{figure}

\begin{figure}
\centering{%
\includegraphics{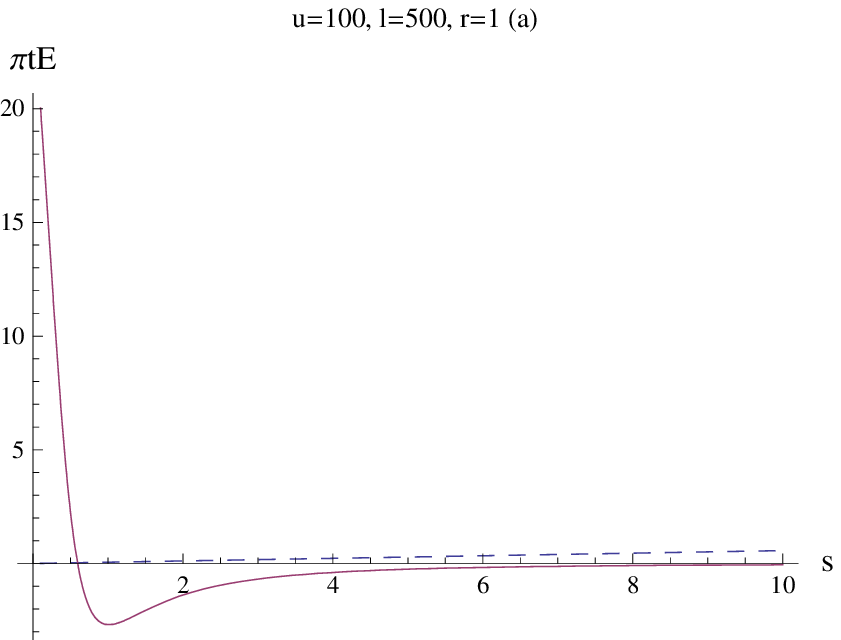}\hskip1cm\includegraphics{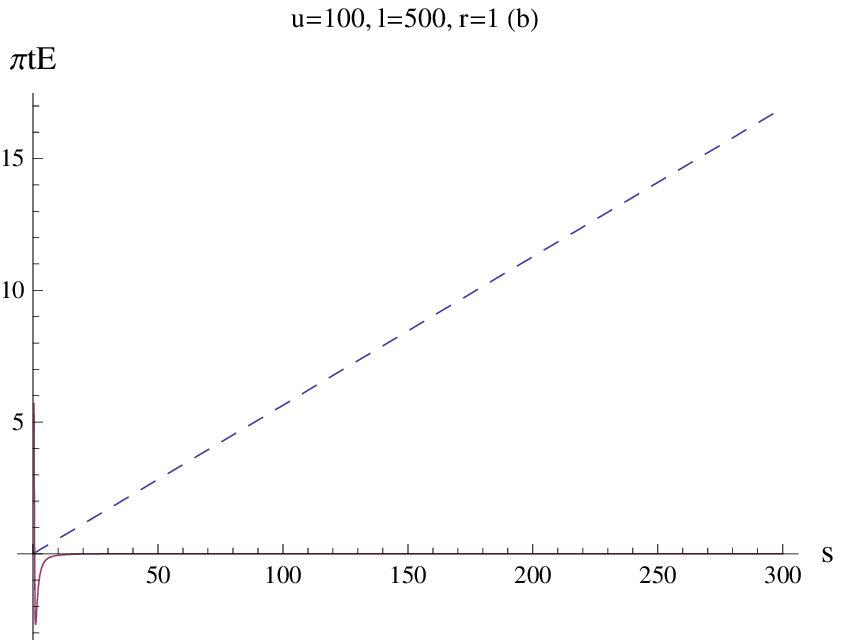}}
 \caption{Graphs of linear (dashed) and nonlinear (solid) parts of 
$\pi tE(s)$ for $r=1$, $u=100$, $l=500$.  (a) Small $s$; linear terms 
are negligible. 
  (b) Large $s$; linear terms dominate and create an attractive 
  force.}
\label{fig:pullgraphs} \end{figure}

\begin{figure}
\centering{%
\includegraphics{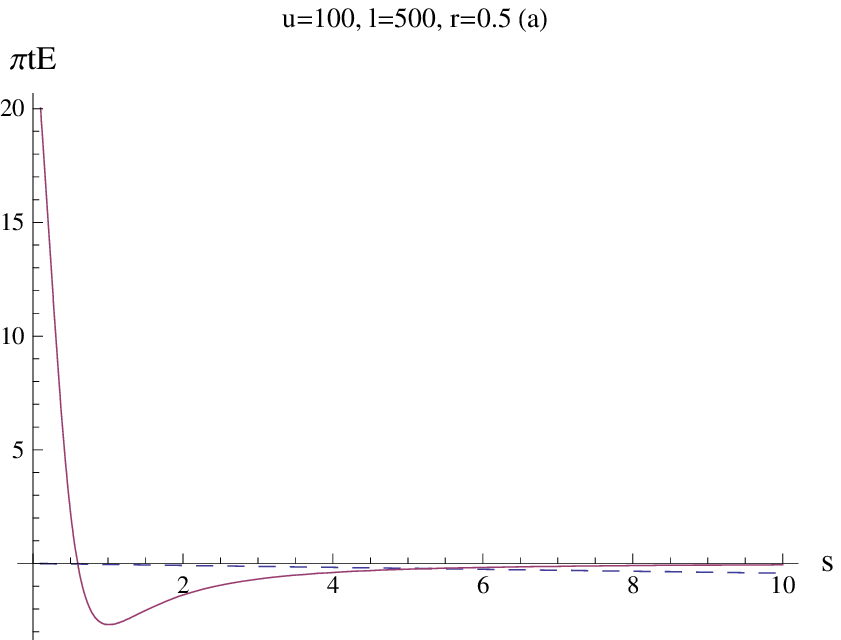}\hskip1cm\includegraphics{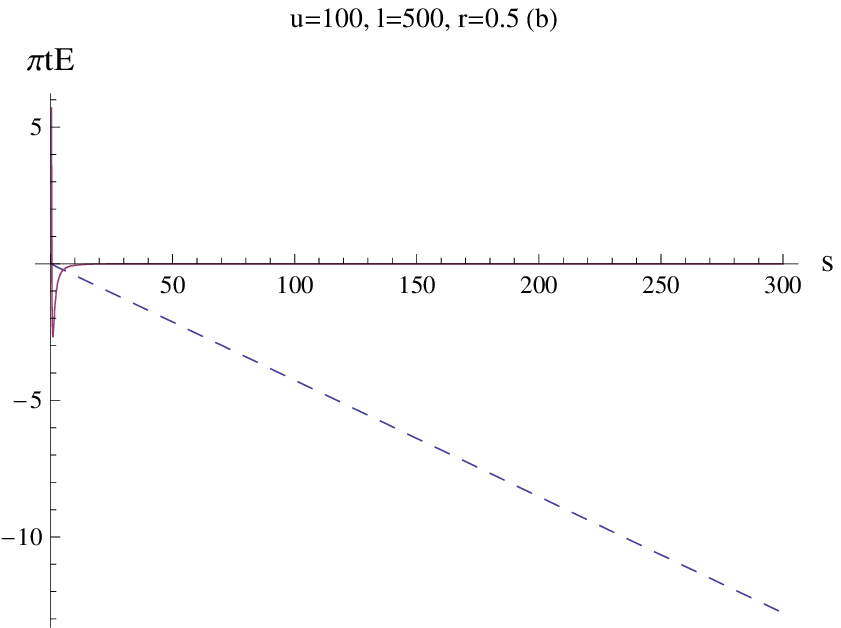}}
 \caption{Graphs of linear (dashed) and nonlinear (solid)
 parts of $\pi tE(s)$ for $r=0.5$, 
$u=100$, $l=500$.  (a) Small $s$; linear terms are negligible. 
  (b) Large $s$; linear terms dominate and create a repulsive
  force.}
\label{fig:pushgraphs} \end{figure}

 Let 
 $c = rt$,  $a = st$, $b=ut$,  $d = L-a =(l-s)t$ (see 
Fig.~\ref{fig:pistoldimenless}).
 We want to examine $E$ as a function of~$s$,
 with $r$ of order unity 
and  $s$, $u$, $l-s$ much larger.
 From (\ref{Etot}) and (\ref{regperenergy}) we have
  \begin{eqnarray} E&=& E_\mathrm{PV} + E_\mathrm{PH} + E_\mathrm{PD} 
 + E_\mathrm{VD} + E_{\mathrm{P}'} \nonumber \\
 &\equiv& \frac{us}{\pi t}\sum_{k=1}^\infty 
 \frac{1-2k^2u^2}{(1+4k^2u^2)^{5/2}} 
+ \frac{us}{\pi t}\sum_{j=1}^\infty 
 \frac{1-2j^2s^2}{(1+4j^2s^2)^{5/2}}
  \nonumber \\
&&{}+\frac{2us}{\pi t} \sum_{j=1}^\infty\sum_{k=1}^\infty
 \frac{1 - 2j^2s^2 - 2k^2u^2 }{ (1+ 4j^2s^2 + 4 k^2 u^2)^{5/2}} 
\nonumber \\
&&{}+ \frac s{2\pi t}   \sum_{j=1}^\infty 
\frac{-1+4j^2s^2}{(1+4j^2s^2)^2 } 
+\frac{2r(l-s)} {\pi t} \sum_{k=1}^\infty
 \frac{1 - 2k^2r^2}{(1 + 4k^2r^2)^{5/2} }\,.
\label{Etotdimenless} \end{eqnarray}
 Let $E_{\mathrm{P}''}$ denote the part of $E_{\mathrm{P}'}$
 proportional to~$s$.  The other term in $E_{\mathrm{P}'}$ 
 (proportional to~$l$) is independent of~$s$ and hence shall be 
ignored in further discussion of the force on the bullet
 (including Figs.\ \ref{fig:pullgraphs}--\ref{fig:pushgraphs}).

 The terms $E_{\mathrm{PV}}$ and $E_{\mathrm{P}''}$ are linear 
functions of~$s$, while the other three terms are nonlinear.
The linear terms dominate the force at large~$s$, and
   the main point of interest is the  confrontation of 
   $E_{\mathrm{P}''}$
 (Casimir energy in the gap) with $E_{\mathrm{PV}}$
  (identified in Sec.~\ref{ssec:force} as the source of the Lukosz
 repulsive force; it is the term that would give an attractive 
Casimir force between the upper and lower walls of the chamber if 
those were allowed to move). 
We shall see that generically the P$''$ term is dominant.
At small~$s$ the nonlinear terms 
dominate and collectively give a function qualitatively similar to 
that in Fig.~\ref{fig:plates2d}.
 Two cases are exhibited in Figs.\ \ref{fig:pullgraphs} 
and~\ref{fig:pushgraphs}.

 In more detail, by approximating the sums by integrals one can 
show that 
 (when the dimensions other than~$r$ are~$\gg 1$) 
 \begin{equation}
 \pi t E_\mathrm{PH} \sim -\,\frac{\zeta(3) u}{16s^2} 
\quad\Rightarrow\quad
\hbox{attractive force}\sim - \,\frac {C'u}{s^3}\,; 
 \label{PHasy}\end{equation} 
 \begin{equation}
 \pi t E_\mathrm{VD} \sim +\, \frac{\zeta(2)}{8s} 
\quad\Rightarrow\quad
\hbox{repulsive force}\sim +  \,\frac C{s^2}\,; 
 \label{VDasy}\end{equation} 
 \[\pi t E_\mathrm{PD} \sim h\left(\frac su\right)\frac1u
 \quad\hbox{for some function $h$,  such that} \]
  \begin{equation}
 s\gg u\gg 1 \;\Rightarrow\; 
 \pi tE_\mathrm{PD} \sim -\,\frac{\zeta(2)}{8s}
\quad\Rightarrow\quad
\hbox{attractive force}\sim -  \,\frac C{s^2} 
 \label{PDasys}\end{equation} 
 so that $E_\mathrm{VD}$ and $E_\mathrm{PD}$ cancel to leading 
order in $1/s$,
and
  \begin{equation}
 u\gg s\gg 1 \;\Rightarrow\; 
 \pi tE_\mathrm{PD} \sim -\,\frac{\zeta(2)}{8u} 
 +O\left(\frac s{u^2}\right)
 \label{PDasyu}\end{equation}
so that the PD force vanishes to leading order in $1/u$.
In the regime $u\gg s$ the PH force dominates the other 
nonlinear terms.
(In the absence of fine tuning, it is still smaller than the 
linear term unless $u\gg s^3$.)
 In the small-$t$ limit (i.e., $s\gg1$)
  the PH force reduces to the first term in 
 (\ref{fsmalla}), which
  is simply the standard Casimir 
force between the left chamber wall and the bullet,
whereas for general $s$  its energy is 
 $E_\mathrm{PH} = uF(s)/\pi t$ --- exactly proportional to the 
parallel-plate function in Fig.~\ref{fig:plates2d}. 
 Finally, for large~$s$ ($s\gg u^{1/3}$) all the nonlinear forces 
are small compared to 
the linear ones, unless the latter happen to cancel. 

 Indeed, the forces arising from the linear terms are
 \begin{equation}
 \pi t^2 F_\mathrm{PV}= -F(u), \qquad
\pi t^2 F_{\mathrm{P}''}= +2F(r), 
 \label{linforce}\end{equation}
 where $F$ is defined by (\ref{platedimenless}).
  Recall that $F$ has a zero at $r_0\approx 0.6$ and a minimum at 
$r_1\approx 1$ and rapidly approaches $0$ at large~$r$.
When $s \gg u^{1/3}$, $F_{\mathrm{P}''}$ exceeds all nonlinear 
forces, in particular the PH force 
(which, we have seen, dominates the  nonlinear forces if 
$s\ll u$).
  Whether the total force is attractive or repulsive at large~$s$ is 
determined by the relative size of the two constant forces in 
(\ref{linforce}),
 and hence on the value of~$r$, $F(u)$ being small and negative.
(1) For $r_0<r\ll u$ and $r$ not too close to $r_0\,$,
 $F_{\mathrm{P}''}$ dominates and the total force is attractive.
 This is the regime in which the cutoff model seems most 
trustworthy physically.
 In particular, it contains the point $r_1\,$,  which one might 
regard as the most ``natural'' value, corresponding to two blocks 
of material in relaxed contact, their effective surfaces separated 
by the typical interatomic spacing.
 (2) For $0\le r< r_0$ and $r$ not too close to $r_0\,$,
 $F_{\mathrm{P}''}$ again dominates and the total force is repulsive.
 In particular, if $r=0$ the two gap forces (P$''$ and VP$'$)
 cancel and the force can be attributed to the negative surface 
energy in the region $\alpha$ (Fig.~\ref{fig:alphabeta}) and the similar 
region next to the part of the bullet outside the barrel.
(3) If $r=r_0\,$, $F_{\mathrm{P}''}$ vanishes and the long-range 
force is purely     $F_\mathrm{PV}$, the repulsive Lukosz force.
 This result is what the piston model was designed to achieve ---
a gedankenexperiment showing that the Lukosz result has, at least 
in principle, some physical reality.
 Unfortunately, that result is attainable only by fine-tuning 
and, moreover,
by pushing $r$ into a regime where the physical relevance of the 
cutoff model is questionable.
(4) For a special value close to $r_0\,$, namely                          
 \begin{equation}
 r  \approx r_0 + \frac{F(u)}{2F'(r_0)} =
r_0 - \frac{\zeta(3)}{32u^2F'(r_0)}   \,,
 \label{zeropt}\end{equation}
 the long-range linear force vanishes.
In this scenario the force is the sum of the PH, PD, and VD terms;
 in the small-cutoff regime of interest,  it is approximately the 
piston force~(\ref{pistforce}). 
This force is always attractive, but exponentially weak at large~$s$. 

In short, the gap plays a spoiler role somewhat like that of the 
outer shaft in the piston model.  In the piston, where shaft 
and chamber have the same width, the shaft force 
precisely cancels the related repulsive force from the VP 
paths in the chamber, leaving the Casimir-like
 force~(\ref{pistforce}). 
The same occurs for the pistol in scenario~4, but in the more 
plausible scenario,~1, the (attractive, $a$-independent) gap 
force overwhelms the interior VP force, precisely because the gap 
 is narrower than the chamber.
In any case, the force arising from outside the chamber depends,
not surprisingly, on the geometrical configuration outside the
chamber, while the force arising inside is fixed by the geometry
of the chamber.

    For the reason mentioned in connection with parallel plates, 
the fact that the gap force becomes repulsive at all at finite 
gap size may be an artifact of the cutoff model.  What one can
say is that the even cruder model of perfect reflection also
displays an artifact, in the form of an attractive force that
diverges as the gap size approaches zero.  In a more realistic
model taking into account the interactions related to condensed
matter physics one would expect the gap force to be reduced, if
never reversed.

 \section{Conclusions} \label{sec:concl}

We have presented a thorough analysis of the vacuum expectation 
value of the stress-energy-momentum tensor in a rectangle.
 The calculational methods involve an exponential ultraviolet 
cutoff and a sum over images (or closed reflecting paths).
 Here we have treated a two-dimensional scalar field; the extension 
to three dimensions and electromagnetism is straightforward and 
under way. Formulas are presented for all tensor components, 
$T_{\mu\nu}(\mathbf{r})$, 
for arbitrary combinations of Dirichlet and 
Neumann boundaries, arbitrary values of the curvature coupling 
$\xi$, and arbitrary values of the cutoff parameter, including 
the limit where the cutoff is removed.
 Forces (which are independent of $\xi$) have been consistently 
calculated both by differentiating energy and by integrating 
pressure. 

Studying the local energy density and stresses (rather than just 
total energy), using a physically motivated ultraviolet cutoff 
(rather than an ``analytic'' regularization scheme), and 
 studying separately the contributions from various classes of 
specularly reflecting paths all help to interpret the physics,
 especially the roles of boundaries and corners.
  Within a cutoff framework one has a clear and consistent 
definition of energy densities and forces. 
 When  different configurations of rigid 
bodies are compared and all contributions (from inside and outside) 
are included, one always finds a cancellation of the 
energy divergences 
and hence an unambiguous force in the  limit of no cutoff.
  The decomposition by paths helps one to understand better the 
cancellations of divergent terms and often to understand 
intuitively the sign of the Casimir force.
Most strikingly, the force on one side of the rectangle includes 
important repulsive components associated with paths 
 \emph{parallel}  to that side:  a divergent term from short paths 
that reflect from the perpendicular sides, and a finite, constant 
term from periodic paths between the two perpendicular sides.
 In piston geometries these forces are cancelled by counterpart 
terms from the exterior of the rectangle, but in more general 
circumstances the problem of their physical interpretation must be 
taken seriously.

 In the later sections of the paper we discuss geometries in which 
the vacuum forces from inside a rectangle might be rigorously 
exhibited.   The box with a loose lid (Fig.~\ref{fig:looselid})
 is closest to what one wants 
to understand, but accurate calculation of the external edge and,
especially, corner
effects remains impractical for now (at least, beyond the scope of 
the present paper).
 The piston model (Fig.~\ref{fig:piston}) 
 studied by previous authors is rigorous and 
exact, but it obscures the point at issue by adding an external 
shaft.
 Our attempt to compromise these two scenarios is the \emph{pistol}
 (Fig.~\ref{fig:pistol}), which unfortunately did not yield a 
robust result. The force on the pistol depends sensitively on the 
cutoff length, as compared to the width of the gap 
 between the bullet and the barrel.
 The only regime in which our quantitative analysis 
(extrapolated to 3D electromagnetism) can be regarded as physically 
trustworthy is that where the gap is small but still larger than 
the cutoff; there the behavior is cutoff-independent but 
the force is attractive.
 Scenarios where the net force is repulsive (in 
particular, one where the gap force vanishes) do exist, but require 
entering the regime where the calculations cannot be taken 
seriously on a quantitative level because one does not know what 
the correct ultraviolet cutoff behavior is
(and because stiction and friction are likely to be the dominant 
effects there); furthermore, making the 
gap force zero or small requires fine tuning within this regime.
 Nevertheless, although no quantitative claims can be made for our 
model (pistol + cutoff) in that regime, we do submit that the model 
is closer to the physical truth than either a model without cutoff 
 (which would predict infinite energies)
  or an analytic regularization that hides the divergences from the 
beginning.
Furthermore, while the repulsive Lukosz component of the force is
robust, the force opposing it is dependent on the scenario
considered (e.g., piston vs.\ pistol, or wide gap vs.\ narrow)
and could in principle be controlled to demonstrate the reality
of the Lukosz force, even if the \emph{net} force is attractive
in all practical experiments.


\ack 
 We thank Martin Schaden and Carlos Villarreal for discussions 
and for 
providing manuscript copies of their unpublished works.
 We thank Gabriel Barton for correspondence and
    Jef Wagner,  Prachi Parashar,
Chris Pope, and Wayne Saslow for useful remarks.
 The numerical plots in Sec.~\ref{sec:pistol-c} were created with 
{\sl Mathematica}, the other graphics with \PiCTeX.

      {\tolerance 8000
This research is supported by the linked NSF Grants PHY-0554849 
(TAMU) and 
 PHY-0554926 (OU) and forms part of our continuing 
collaboration with Ricardo Estrada. \hfil
K.~A.~Milton also received support from DOE Grant
\hbox{DE-FG02}-04ER41305.
L.~Kaplan is supported by NSF Grant PHY-0545390.
    K.~Kirsten is supported by NSF Grant PHY-0757791.
\par}

\end{document}